\documentclass{aa}
\usepackage{graphicx}
\usepackage{txfonts}
\usepackage{lipsum}
\usepackage{lscape}
\usepackage{placeins}  
\usepackage{graphicx}
\usepackage{epstopdf}
\usepackage{amsmath}
\usepackage{amssymb}
\usepackage{appendix}
\usepackage{subfigure}
\usepackage{stfloats}
\usepackage{CJKutf8}
\usepackage{diagbox}
\usepackage{threeparttable}
\usepackage{booktabs}
\usepackage{bm}
\usepackage{gensymb}
\usepackage{array}
\usepackage{xcolor}
\usepackage{hyperref}

\begin{document}

\title{Radiative cooling effects on black hole hot accretion flows around the sub-Bondi radius}
   
\author{Mu-Qing Liu\inst{1}
    \and Xiao-Hong Yang\inst{1}\thanks{Corresponding author: yangxh@cqu.edu.cn(XH)}
    \and De-Fu Bu\inst{2}\thanks{Corresponding author: dfbu@shnu.edu.cn(DF)}
    }
\authorrunning{Liu, Yang, Bu}

\institute{Department of Physics and Chongqing Key Laboratory for Strongly Coupled Physics, Chongqing University, Chongqing 400044, People's Republic of China
	\and Shanghai Key Lab for Astrophysics, Shanghai Normal University, 100 Guilin Road, Shanghai 200234, China}

\date{Received September 30, 20XX}

 
  \abstract
   {}
   {It is difficult to implement numerical simulations on a region extending from the vicinity of a black hole to the Bondi radius. Most previous numerical simulations have primarily concentrated on the region close to the black hole. They found that strong winds can be generated in the hot accretion flows near the black hole, and that radiative cooling significantly affects the strength of these winds. However, the effects of radiative cooling on the production and properties of winds around the Bondi radius remain unclear.} 
   {In this paper, we perform two-dimensional magnetohydrodynamic simulations to study the impact of radiative cooling on the dynamics and wind production in hot accretion flows around the sub-Bondi radius.}
   {As the increase of mass accretion rate, radiative cooling gradually becomes strong, resulting in a reduction in the thickness of the accretion disk (defined as the accretion flows within the density scale height). Based on the H{\o}iland criterion, we find that within the accretion disk, the region of convective stability accounts for $\sim$55--62 \%, and therefore the accretion flows are marginally stable in convective stability. In the runs with weak radiative cooling, the winds play a significant role in the inward decrease of the mass inflow rate. In the runs with strong radiative cooling, the mass outflow rate of winds is significantly reduced and then the inward decrease of the mass inflow rate is mainly attributed to turbulence driven by magnetorotational instability and convection. Radiative cooling has the potential to suppress accretion processes and reduce the power of winds.}
   {}

   \keywords{accretion --
   			 accretion discs -- black hole physics --
   			 hydrodynamic
               }
   \maketitle
   \nolinenumbers

\section{Introduction}
Supermassive black holes (SMBHs) are widely believed to exist at the centers of galaxies. In nearby galaxies, most of active galactic nuclei (AGNs) are low-luminosity AGNs (LLAGNs) with luminosities ranging from $10^{-9}$ to $10^{-1} L_{\rm Edd}$, where $L_{\rm Edd}$ is  the Eddington luminosity \citep{Ho_2008, Ho_2009}. Hot accretion flows around SMBHs are generally employed to explain the observed properties of LLAGNs. Hot accretion flows are characterized by lower density and higher temperature, which causes them to be radiatively inefficient. The energy dissipated in hot accretion flows is advected into SMBHs along with the accreting matter. These flows are commonly referred to as advection-dominated accretion flows (ADAFs, \citealt{Ichimaru_1977, Rees_1982, Narayan_1994, Narayan_1995, Abramowicz_1995}).

In recent years, a series of numerical studies have investigated the hot accretion flows in detail \citep{Stone_1999,Igumenshchev_1999, Igumenshchev_2000, Hawley_2001, Stone_2001, Hawley_2002, Yuan_2012}. One of the significant findings is the identification of strong winds launched from the radiatively-inefficient hot accretion flows \citep{Yuan_2012, Yuan_2015}. \cite{Stone_1999} implemented the two-dimensional hydrodynamic (HD) simulations of non-radiative hot accretion flows and found that the mass inflow rate diminishes as the radius decreases. \cite{Yuan_2012} confirmed that the wind-induced mass loss results in a reduction of the mass inflow rate as the accretion gas moves inward. Based on observations from the Chandra X-ray Observatory, \cite{Wang_2013} confirmed the existence of winds originating from hot accretion flows surrounding the SMBH in our galactic center (Sgr A$^{*}$).

For further investigating the properties of winds, \cite{Yuan_2015} performed the trajectory analysis of the virtual Lagrangian particles based on a 3D general relativistic magnetohydrodynamic (MHD) simulation of non-radiative hot accretion flows around a Schwarzschild BH. They found that, at small scale ($r<10^3 r_{\rm s}$, with $r_{\rm s}$ being Schwarzschild radius), the radial profile of the mass flux of winds can be approximately expressed as $\dot{M}_{\rm wind}\approx\dot{M}_{\rm BH} (r/20r_{\rm s})$, where $\dot{M}_{\rm BH}$ is the mass accretion rate at the BH horizon. This result suggests that most of the winds are produced at large radii within the region of $r<10^3 r_{\rm s}$. When radiative cooling is considered, the strength of winds depends on the accretion rate \citep{Bu_2018}. Based on the 2D HD simulations, \cite{Bu_2018} found that radiative cooling becomes important with the increase of accretion rate. When the accretion rate increases to a value, winds become very weak. Their study did not consider the effect of magnetic field. However, if magnetic field is taken into account, the wind strength may be changed due to the presence of the magnetic-driving winds \citep{Blandford_1982, Cao_2011, Li_2014}. The aforementioned studies primarily investigate the generation of winds on small scales ($r<10^3 r_{\rm s}$), i.e. on the region close to the BH. Therefore, the important questions are whether or not the winds can be produced at large scale ($r>10^3 r_{\rm s}$) and how the accretion rate influences the wind strength.

The accretion processes around a BH are generally believed to extend out to the Bondi radius \citep{Bondi_1952}. For the spherically symmetric flow without rotation and radiation, within the Bondi radius, the gas is gravitationally bound. For a realistic accretion flow, due to the presence of gas angular momentum and radiative cooling, the accretion flow can not be adequately described by the Bondi solution. A dynamical range extending from the Bondi radius to the vicinity of a BH is excessively large. It is difficult to implement numerical simulation on this dynamical range. A feasible approach is to independently implement numerical simulations at small scales (the vicinity of a BH) and large scales (near the Bondi radius) to investigate the dynamics of accretion flows. Several previous works have implemented HD and MHD simulations to study the hot accretion flows at large scales \citep{Li_2013, Bu_2016a, Bu_2016b, Inayoshi_2018, Inayoshi_2019, Mosallanezhad_2022}. \cite{Bu_2016a, Bu_2016b} and \cite{Inayoshi_2018} studied the radiatively-inefficient accretion flows around the sub-Bondi radius. \citet{Li_2013} implemented HD simulations to study the effect of radiative cooling on the accretion flows around the sub-Bondi radius. They found that strong radiative cooling promotes accretion. In the HD simulations, the $\alpha$-viscosity prescription is usually employed to mimic the anomalous shear stresses for angular momentum transport and the heating effects resulting from the work done by stresses. However, HD simulation results depends on the assumption of $\alpha$. In reality, the transport of angular momentum in accretion flows should be driven by Maxwell stresses arising from MHD turbulence, which is induced by the magnetorotational instability (MRI; \citealt{Balbus_1998}). Therefore, it is necessary to perform MHD simulations on the accretion flows around the Bondi radius.

In this paper, we perform MHD simulations to investigate the effects of radiative cooling on the dynamics and wind production in hot accretion flows around the sub-Bondi radius. In Section 2, we will describe the method and model setup. In Section 3, we will will analyze the main results. In Sections 4 and 5, we will present a discussion and a summary of the results, respectively.
\section{Method}
\subsection{Basic equations and numerical method}

To describe the BH hot accretion flows around the sub-Bondi radius, we employ the MHD equations as follows:
\begin{equation}
	\frac{\partial\rho}{\partial t} + \nabla\cdot{(\rho \bm{v})} = 0,
	\label{eq1}
\end{equation}	

\begin{equation}
	\frac{\partial (\rho \bm{v}) }{\partial t} + \nabla\cdot{ [ \rho\bm{v}\bm{v} - \bm{B}\bm{B} + (p_{\rm g} + \frac{\bm{B}^2}{2}) ] } = -\rho\nabla{\Psi},
	\label{eq2}
\end{equation}	

\begin{equation}
	\frac{\partial E_{\rm t}}{\partial t} + \nabla \cdot {[ (E_{\rm t} + p_{\rm g})\bm{v} - \bm{B}(\bm{v} \cdot \bm{B}) ]} = -\rho \bm{v} \cdot \nabla \Psi - Q^{-}_{\rm brem},
	\label{eq3}
\end{equation}	

\begin{equation}
	\frac{\partial \bm{B} }{\partial t} - \nabla\times{(\bm{v}\times\bm{B})} = 0.
	\label{eq4}
\end{equation}	
Here, $\rho$, $p_{\rm g}$, $\bm{v}$, and $\bm{B}$\footnote{A factor of $1/\sqrt{4\pi}$ has been absorbed.} are the density, pressure, velocity, and magnetic field, respectively. $E_{\rm t}$ is the total energy density, $E_{\rm t} = e + \rho\bm{v}^2/2 + \bm{B}^2/2 $, where $e$ is the internal energy. We adopt an equation of state of ideal gases, $p_{\rm g} \equiv(\gamma - 1)e\equiv\rho \kappa_{\rm B} T_{\rm g} /\mu m_{\rm p}$, and set $\gamma$ = 5/3, where $\mu$, $\kappa_{\rm B}$, $m_{\rm p}$, and $T_{\rm g}$ are the mean molecular weight, the Boltzmann constant, the proton mass, and the gas temperature, respectively. In this paper, we set $\mu=0.6$. The gravitational potential ($\Psi$) of the center BH is described using the pseudo-Newtonian potential introduced by \cite{Paczy_1980},  i.e. $\Psi(r) = -GM/(r-r_{\rm s})$, where $G$ is the gravitational constant, $M$ is the mass of the central BH, and $r_{\rm s} \equiv 2GM/c^2$ is the Schwarzschild radius. ${Q}^-_{\rm brem}$ is the bremsstrahlung cooling term given by \cite{Li_2013}:
\begin{equation}
	{Q}^-_{\rm brem} = \alpha_f r_e^2 m_e c^3 n^2 (32/3)(2/\pi)^{1/2}(\frac{\kappa_{\rm B}T_{\rm g}}{m_ec^2})^{1/2},
	\label{eq_Q-}
\end{equation}
where $\alpha_f$ is the fine structure constant, $r_e$ is the classical electron radius, $c$ is the speed of light, and $n$ is the number density. 

We set the BH mass to be $M=10^8 M_{\odot}$. Assuming the axisymmetric condition, i.e $\partial/\partial\phi = 0$, we employ the PLUTO code to numerically solve Equations (\ref{eq1})--(\ref{eq4}) in a spherical coordinates ($r$, $\theta$,$\phi$). The PLUTO code is a Godunov-type code solving the HD and MHD equations \citep{Mignone_2007, Mignone_2012}. The bremsstrahlung cooling term in equation (\ref{eq3}) is considered as a source term, which is solved implicitly to update the gas temperature and internal energy in the PLUTO code.

The computation domain is from $r_{\rm in} = 2 \times 10^{-2}r_{\rm B}$ to $r_{\rm out} = 5r_{\rm B}$ in radial direction and from $\theta_1 = 0$ to $\theta_2 = \pi$ in angular direction. Here, $r_{\rm B}$ is the Bondi radius, which is determined based on the gas temperature at infinity $T_{\infty}$. We set $T_{\infty}\sim 1.7\times 10^6 \rm K$, and then $r_{\rm B}$ equals $10^6 r_{\rm s}$. In the radial direction, we employ $256$ zones with the radial size ratio of $(\triangle r)_{i+1}/(\triangle r)_{i}=1.0177$. In the angular direction, 104 zones with the angular size ratio of $(\triangle \theta)_{j+1}/(\triangle \theta)_{j}=0.98106$ are distributed over $0<\theta<\pi/2$ and 104 zones with the angular size ratio of $(\triangle \theta)_{j+1}/(\triangle \theta)_{j}=1.01930$ are distributed over $\pi/2<\theta<\pi$. The smallest angular size is $\triangle \theta=0.00480956$ at $\theta=\pi/2$. Axisymmetric boundary conditions are employed at the poles and outflow boundary conditions are employed at the inner and outer radial boundaries.

\subsection{Initial condition}

Initially, we assume that a torus in rotating equilibrium with constant specific angular momentum is embedded in a non-rotating, low-density medium. We follow \cite{Kato_2004} to determine properties of the torus. The specific angular momentum of the torus is set to be $l = (GMr_0^3)^{1/2} / (r_0-r_{\rm s})$, where $r_0$ (=  $r_{\rm B}$) is the radius of the torus center. For the initial torus, a polytropic relation between pressure ($p_{\rm t}$) and density ($\rho_{\rm t}$) is $p_{\rm t}=K \rho_{\rm t}^{1+1/n}$, where $K$ is a constant and $n$ is set to be equal to 3. The initial density and pressure distributions of the torus are given by
\begin{equation}
	\rho_t = \rho_0(1-\frac{\gamma}{c^2_{\rm s,0}}\frac{\widetilde{\Psi}-\widetilde{\Psi}_0}{n+1})^n,
	\label{rho_t}
\end{equation}
and 	
\begin{equation}
	p_t =
	\rho_0 \frac{c^2_{\rm s,0}}{\gamma}(\frac{\rho_t}{\rho_0})^{1+\frac{1}{n}},
	\label{p_t}
\end{equation}
where $c^2_{\rm s,0}$ is the sound speed at the torus center, $\widetilde{\Psi}$ ($\equiv \Psi + (l/r\sin{\theta})^2/2$) is the effective potential, and $\widetilde{\Psi}_0$ is the effective potential at torus center, respectively. We can give the gas temperature ($T_{\rm g,0}$) at the torus center to determine $c^2_{\rm s,0}$ ($\equiv\gamma k_{\rm B} T_{0}/\mu m_{\mu}
$) and then determine the initial density and pressure of the torus. Here, we set $T_{\rm g,0}$ = $2\times10^5$ K. For the initial ambient medium, we assume the initial density $\rho_a = 10^{-6}\rho_0(\frac{r_{\rm in}}{r})^4$ and the initial pressure $p_a = \frac{GM\rho_a}{r}$. The density floor is also set to be equal to $\rho_{\rm a}$.

\subsection{Magnetic field setup}

The magnetic field topology at the Bondi radius is observationally unknown. According to \citet{Han_2017}, although the magnetic field may exhibit disorder on small scales, it is expected to be ordered on galactic scales. The introduced weak magnetic field in numerical simulations can grow under MRI perturbations. \citet{Beckwith_2008} investigated the impact of the initial magnetic field topology on accretion flows. They employed three different topological configurations, such as dipole, quadrupole, and toroidal fields. Their results indicate that once the accretion flow reaches a quasi-steady state, the initial topology of the magnetic field does not significantly affect its primary characteristics. \citet{Yuan_2014} also pointed out that the initial topology of the magnetic field exerts minimal influence on the characteristics of the accretion flows once they have developed into a quasi-steady state. However, it may have a certain impact on the properties of the jet. Generally, the dipole field generates a strong and steady jet, whereas the quadrupole field produces a weaker and intermittent jet. In contrast, the toroidal field yields almost no jet at all. Here, we initially choose a vector potential to introduce a weak seed magnetic field.

Similar to the magnetic field configuration proposed by \cite{Penna_2013}, we construct the initial magnetic fields that are purely poloidal, i.e. only the toroidal component ($A_{\phi}$) of magnetic potential is nonzero. The toroidal component is then

\begin{equation}
	\bm{A_{\phi}} \propto \left\{
	\begin{array}{rcl}
		q \ {\rm sin}(\frac{N_{\rm L}\pi( g(r) - g(r_{\rm beg})}{g(r_{\rm end})-g(r_{\rm beg})} )   &&  q > {0.0}\\
		0.0     && \rm{otherelse}\\
	\end{array} \right. ,
	\label{loop}
\end{equation}
where
\begin{equation}
	\left\{
	\begin{aligned}
		g(r) &= r^{2/3}+\frac{15}{8}r^{-2/5} \\
		q &= \rm (cos\theta)^{N_{\rm q}} \rm sin^{3} \theta \max(\rho-\rho_{\rm cut},0)
	\end{aligned}
	\right. .
\end{equation}
Here,  $r_{\rm beg}$ and $r_{\rm end}$ are the inner and outer boundaries of the magnetized region, respectively.  $\rho_{\rm cut}$ is the cut-off density and we set $\rho_{\rm cut} = 10^{-4}\rho_0$. $N_{\rm L}$ is a parameter to control the number of the poloidal magnetic loops along the $r$ direction. $N_{\rm q}$ is a parameter that determines whether a magnetic loop is divided into two mirror-symmetric loops relative to the equator or not. When $N_{\rm q}=0$, only one loop is distributed along the $\theta$ direction. When $N_{\rm q}=1$, a loop is divided into two loops that are mirror symmetric relative to the equator.

\begin{figure}[htb]
	\includegraphics[width=0.5\textwidth]{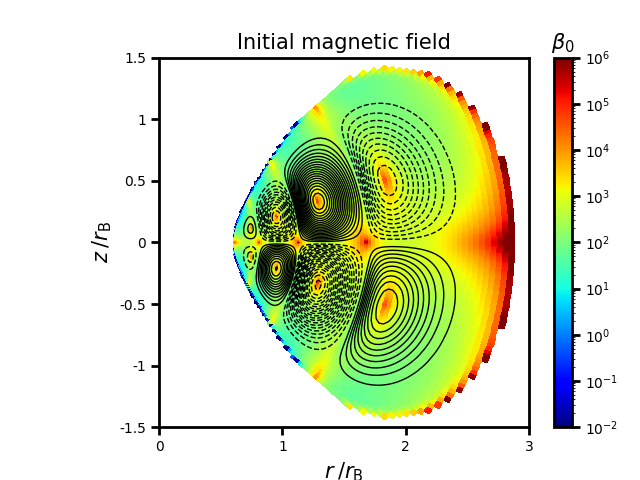}
	\centering \caption{The distribution of the initial magnetic field (lines) and plasma $\beta_0$ (colors). Solid and dashed lines represent clockwise and counterclockwise magnetic field lines, respectively.}
	\label{Fig1_InitMag}
\end{figure}

\begin{figure*}[htb]
\centering
\setlength{\tabcolsep}{1.2pt} 
\begin{tabular}{cccc}
\includegraphics[width=0.5\textwidth]{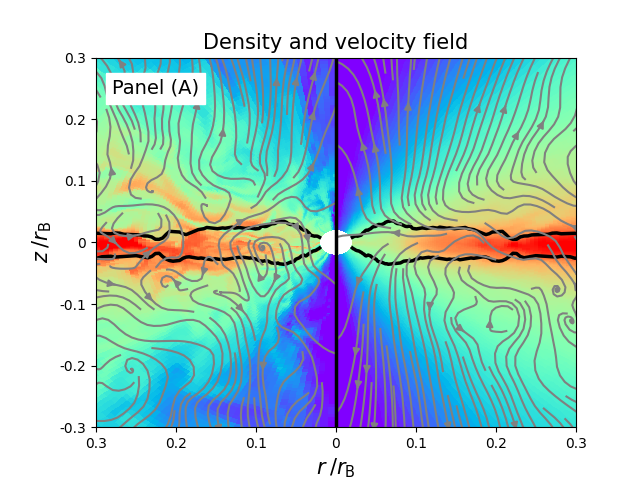} &
\includegraphics[width=0.5\textwidth]{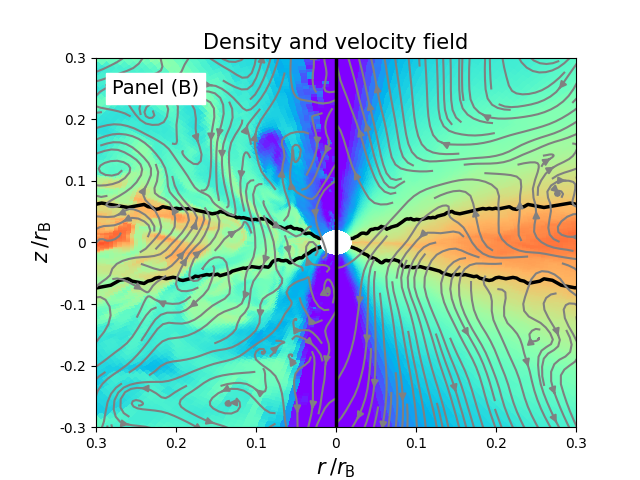} \\
\includegraphics[width=0.5\textwidth]{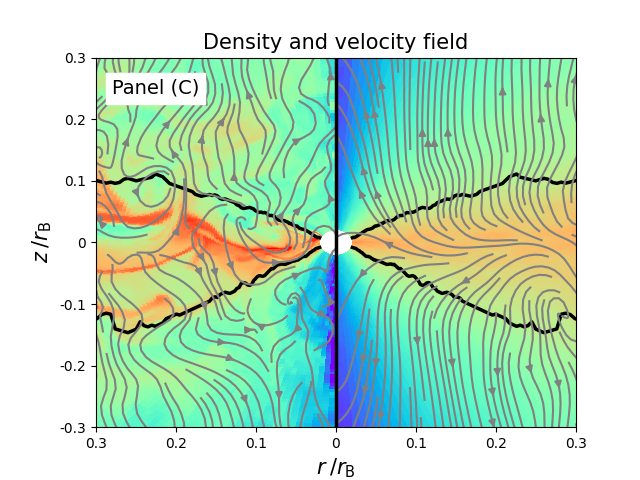} &
\includegraphics[width=0.5\textwidth]{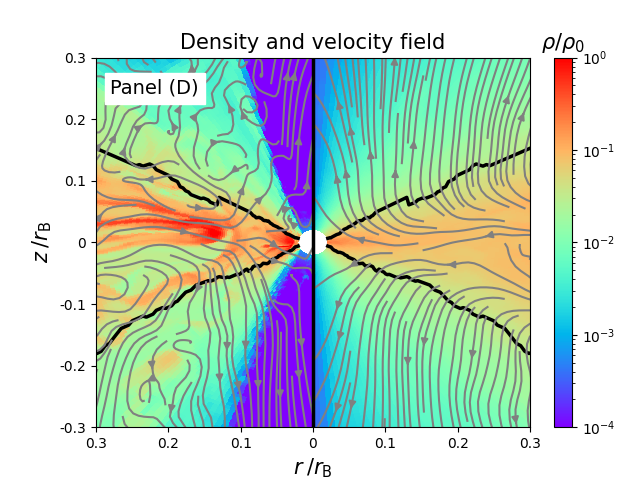}
\end{tabular}
\caption{The distribution of density and velocity field. In each panel, the left side presents a snapshot at $4.0 t_{\rm B}$, while the right side displays the time-averaged distribution. Colors represent the density, measured in units of the density at the center of the initial torus. The streamlines reflect the velocity field on the poloidal plane. The black solid line marks the scale height of accretion flows. The scale height is defined as the height at which the density drops by a factor of $e$ relative to the density at the equatorial plane. Panel (A): Run R22. Panel (B): Run R23. Panel (B): Run R24. Panel (D): Run N24.}
\label{Fig2_Density}
\end{figure*}

For each magnetic loop, we specify the scale factor in equation (\ref{loop}) to determine the magnetic field strength. The scale factor is determined by the ratio ($\beta_{0}=p_{\rm g}/p_{\rm m}$) of the gas pressure to the magnetic pressure ($p_{\rm m}$) at the location of maximum magnetic pressure over each loop. In our runs, we set $N_{\rm L}=4$, $N_{\rm q} = 1$ and $\beta_{0}=1000$. The configuration of magnetic fields are shown in Figure \ref{Fig1_InitMag}.

\begin{table*}
	\caption[]{Main parameters}
	\label{Table1}
	\begin{center}
		\begin{tabular*}{\textwidth}{@{\extracolsep{\fill}}cccccccc}
			\toprule[1.3pt]
			Run name & $\rho_{0}(g\cdot cm^{-3})$ & $\dot{M}_{\rm acc}(\dot{M}_{\rm Edd})$ & $ s_{\rm {in}}$ & $s_{\rm {out}}$ & $p$ & $A(\%)$ & Radiative cooling\\
			\midrule[1.1pt]
			R22  &$10^{-22}$& $1.7 \times 10^{-6}$ & $1.64$ & $1.94$ & $0.05$  & $62.00\%$  & ON  \\
			R23  &$10^{-23}$& $3.9 \times 10^{-7}$ & $1.29$ & $1.68$ & $0.23$  & $58.17\%$  & ON  \\
			R24  &$10^{-24}$& $9.0 \times 10^{-8}$ & $0.86$ & $1.35$ & $0.37$  & $55.60\%$  & ON  \\
			N24  &$10^{-24}$& $1.3 \times 10^{-7}$ & $0.76$ & $1.34$ & $0.50$  & $56.42\%$  & OFF \\
			\bottomrule[1.3pt]
		\end{tabular*}
	\end{center}
	\begin{tablenotes}{}
		\item{Notes. Column (1) means run name. \quad
			Column (2) presents the density at the center of the initial torus. \quad
			Column (3) gives the net accretion rate at inner boundary. \quad
			  Column (4) gives the power-law indices for the inflow, characterized by a radial profile proportional to $r^{s_{\rm in}}$. \quad
			Column (5) gives the power-law indices for the outflow, characterized by a radial profile proportional to $r^{s_{\rm out}}$. \quad
			Column (6) gives the Power-law indices for the density at the equational panel. The density is described by a radial profile that is proportional to $r^{-p}$. \quad
			Column (7) gives the proportion of the area occupied by the convectively stable regions within the accretion disk. \quad
			Column (8) means that radiative cooling was either included or excluded. "ON" means that radiative cooling was implemented during the simulation, while "OFF" indicates that it was not.}
	\end{tablenotes}
\end{table*}

\section{Result}
Table 1 summarized our simulation runs. In our simulations, we can adjust the accretion rate by varying the density ($\rho_{0}$) at the center of the initial torus. As shown in Table 1, the density increases from 10$^{-24}$ g cm$^{-3}$ in run R24 to 10$^{-22}$ g cm$^{-3}$ in run R22. For the purpose of comparison, we do not incorporate radiative cooling in the simulation of run N24. In the simulations of runs R22--R24, we initially let the runs without radiative cooling evolve from the initial state to a quasi-steady state. Subsequently, we incorporate radiative cooling into these runs and then permit them to evolve continuously to achieve a quasi-steady state once again. In general, we incorporate radiative cooling into these runs after $t=2 t_{\rm B}$, where $t_{\rm B}$ is the Keplerian orbit period at $r=r_{\rm B}$. In the following analysis, we perform time-averaging over the time interval of 4--5 $t_{\rm B}$. During this time interval, we gather 400 data files with equal time intervals.  

In magnetized accretion flows, the MRI is the primary mechanism for angular momentum transport. To ensure that the MRI is adequately captured in numerical simulations, it is essential to resolve its fastest growing MRI mode. The characteristic wavelength of the fastest growing MRI mode is given by $\lambda_{\rm MRI} = 2\pi V_A / \Omega$, where $V_A$ is the Alfvén speed and $\Omega$ is the angular velocity. Following \citet{Hawley_2011, Hawley_2013}, we define the MRI quality factors in the radial and polar directions as $Q_r = 2\pi V_{A,r} / \Omega \, \Delta{r}$ and $ Q_\theta = 2\pi V_{A,\theta} / \Omega \, r\Delta{\theta}$, where $V_{A,r}$ and $V_{A,\theta}$ are the radial and polar components of the Alfvén velocity, respectively, and $\Delta {r}$ and $r\Delta{\theta}$ represent the local grid size in the corresponding directions. These quality factors quantify the number of grid cells available to resolve the fastest growing MRI mode. Previous studies have suggested that maintaining $Q_r \gtrsim 10$ and $Q_\theta \gtrsim 10$ is sufficient to sustain MRI-driven turbulence throughout the nonlinear regime \citep{Hawley_2011, Hawley_2013}. In our simulations, within the region that achieves a quasi-steady state, the MRI quality factors exceed the threshold value of 10 throughout the majority of the domain. This demonstrates that our simulations are sufficiently well-resolved to effectively capture MRI turbulence, thereby ensuring the reliability of the derived accretion dynamics and wind properties.

In all the runs, accretion flows achieve a quasi-steady state within the region of $r<\sim 0.3 r_{\rm B}$, where the net accretion rate remains approximately constant (see the dotted line in figure \ref{Fig11_AccretionRate_r}). Therefore, in the following analysis, we focus on the accretion flows within the quasi-steady state region. For the purpose of analysis, we employ two distinct methods for averaging physical quantities. The time-averaged value of a given quantity ($a$) in each grid ($r$, $\theta$) is defined as follows:
\begin{equation}
	 \langle a \rangle_{\rm t} = \frac{ \int_{t_1}^{t_2} a \, dt }{\int_{t_1}^{t_2} \, dt}.
\end{equation} 
The time- and angular averaged value of a given quantity ($a$) at each radius is defined as follows:
\begin{equation}
	\langle \langle a \rangle \rangle = \frac{ \int_{t_1}^{t_2} \,\int_{\theta_1}^{\theta_2}a \,  \sin \theta d\theta \, dt}{\int_{t_1}^{t_2} \,\int_{\theta_1}^{\theta_2}  \sin \theta d\theta \, dt}.
\end{equation} 
Here, the time interval spans from $t_1 = 4 t_{\rm B}$ to $t_2 = 5 t_{\rm B}$. For evaluating the angular averaged value around the equatorial plane, the integration limits for the polar angle are defined as $\theta_1 = 86^\circ$ and $\theta_2 = 94^\circ$.
  
\begin{figure}[htb]
	\includegraphics[width=0.5\textwidth]{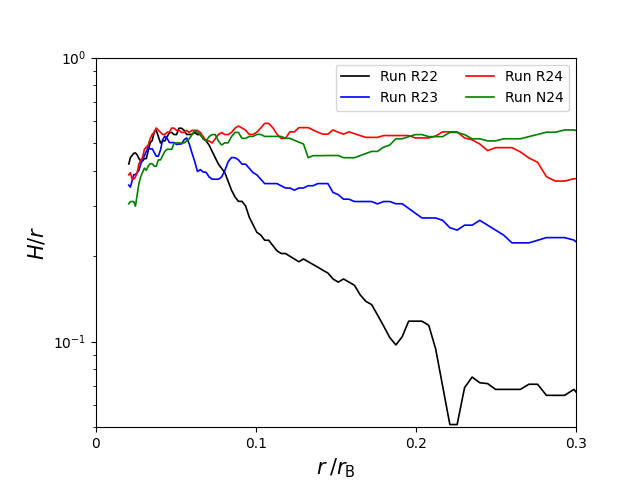}
	\centering \caption{Radial dependence of the ratio ($H/r$) of density scale height to radius. The black, blue, red, and green lines correspond to runs R22, R23, R24, and N24, respectively.}
	\label{Fig3_DiskH}
\end{figure}

\begin{figure}[htb]
	\includegraphics[width=0.5\textwidth]{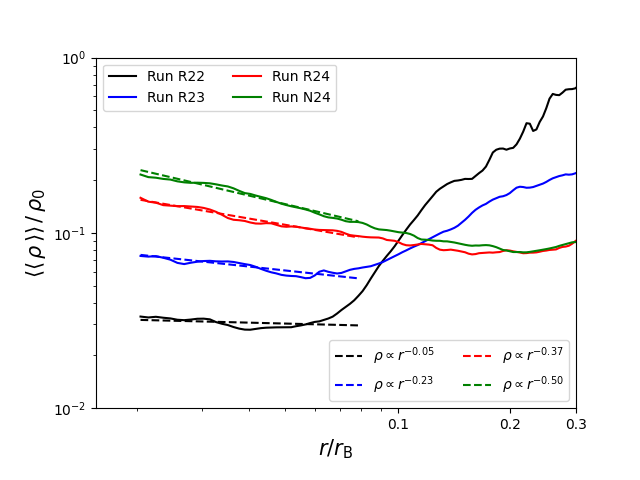}
	\centering \caption{The radial profiles of the time-averaged and angle-averaged density near the equatorial plane. The angle averaging is implemented over the angle range of $84\degree$ to $96\degree$. The black, blue, green, and red lines correspond to runs R22, R23, R24, and N24, respectively.}
	\label{Fig4_Density_r}
\end{figure}

\begin{figure}[htb]
    \centering
    \includegraphics[width=0.5\textwidth]{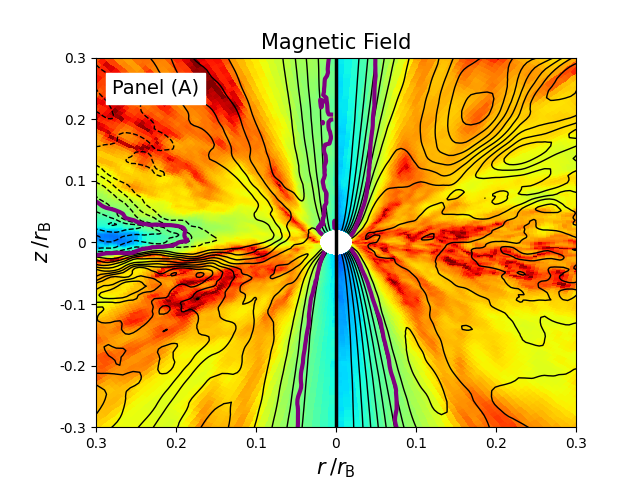}
    \includegraphics[width=0.5\textwidth]{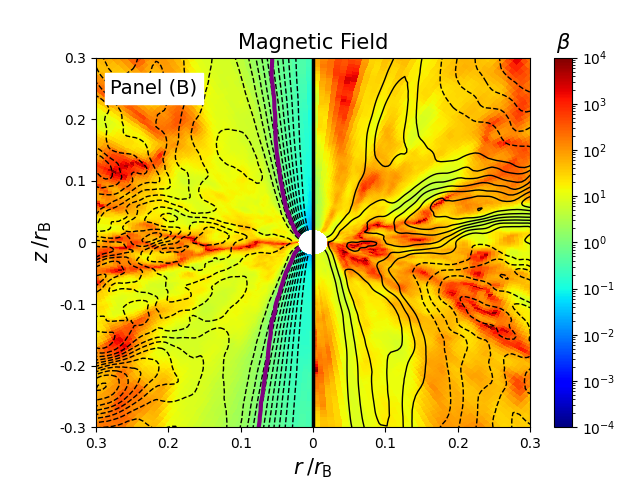}
    \caption{The distribution of the plasma $\beta$ and poloidal magnetic fields. Colors represent the ratio (${\beta}$) of the time-averaged gas pressure to the magnetic pressure of the time-averaged magnetic field. The solid and dashed lines reflect the magnetic fields on the poloidal plane. The purple line indicates the interface at which $\beta = 1$. Panel (A): Runs R22 (left) and R23 (right). Panel (B): Runs R24 (left) and N24 (right).}
    \label{Fig5_Beta}
\end{figure}

\begin{figure}[htb]
    \centering
    \includegraphics[width=0.5\textwidth]{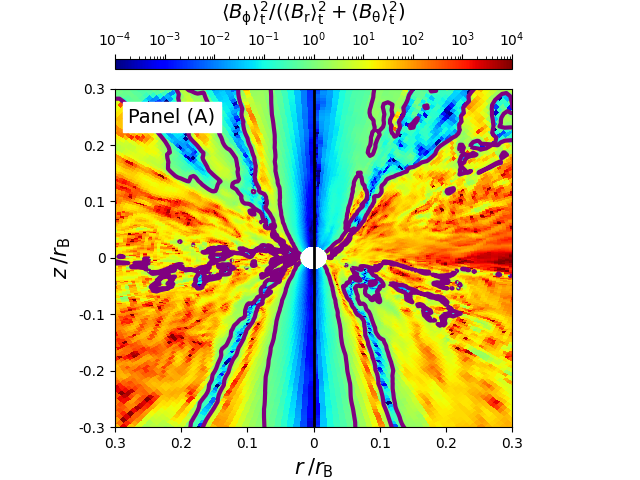}
    \includegraphics[width=0.5\textwidth]{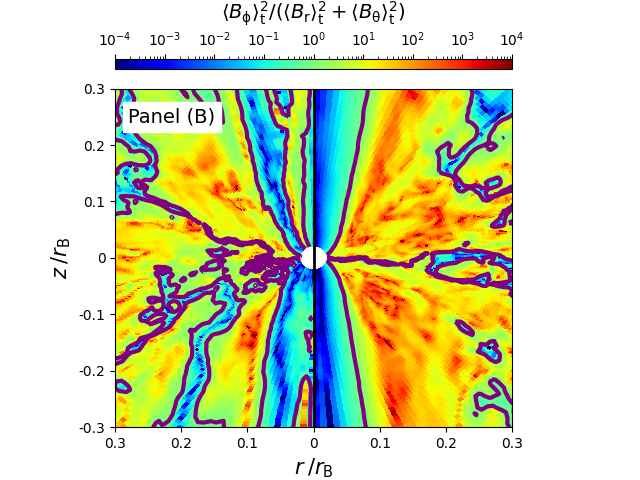}
	\caption{The distribution of the strength ratio between the toroidal and poloidal components of the mean magnetic field. The purple line indicates the interface where the toroidal magnetic field strength equals the poloidal magnetic field strength. Panel (A): Runs R22 (left) and R23 (right). Panel (B): Runs R24 (left) and N24 (right).}
	\label{Fig6}
\end{figure}

\begin{figure}[htb]
	\centering
	\includegraphics[width=0.5\textwidth]{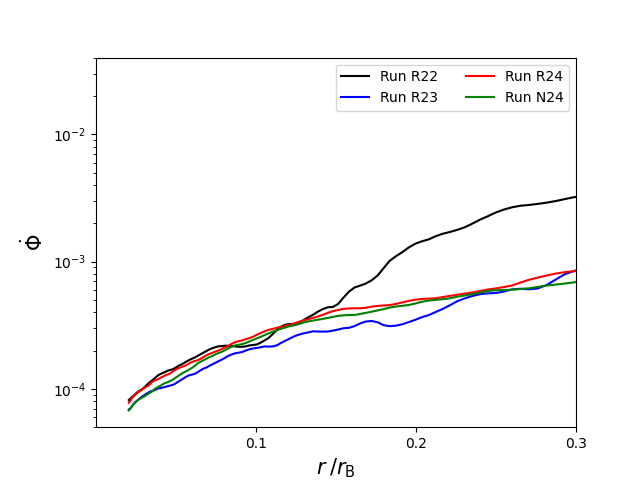}
	\caption{The radial profiles of normalized magnetic flux around equatorial plane. The black, blue, red, and green lines correspond to runs R22, R23, R24, and N24, respectively.}
	\label{Fig7_MagneticFlux}
\end{figure}

\subsection{Structure of accretion flows}

Figure \ref{Fig2_Density} shows the distribution of density and velocity for runs R22--R24 and N24. The black solid line in Figure \ref{Fig2_Density} marks the density scale height ($H$) of accretion flows. The density scale height is defined as the vertical distance at which the density drops by a factor of $e$ relative to the density at the equatorial plane. Here, we define the accretion disk as the accretion flows within the density scale height. From Figure \ref{Fig2_Density}, we can see that outflows are generated as matter escapes from the accretion disk. In order to compare the density scale height of various runs, we plot Figure \ref{Fig3_DiskH}, which shows the radial dependence of the ratio ($H/r$) of density scale height to radius. In run R24 with radiative cooling, the density scale height is comparable to that observed in run N24 without radiative cooling. Both runs exhibit similar values for $H/r$, ranging from approximately 0.3 to 0.6. This indicates that radiative cooling is very weak in R24. In run R23, $H/r$ is slightly lower, approximately $0.2$--$0.5$, due to radiative cooling. In run R22, the density scale height is the lowest among all the runs, and beyond 0.1 $r_{\rm B}$ the $H/r$ value is lower than 0.2, which implies that the accretion flows undergo a significant contraction due to strong radiative cooling. Therefore, as the accretion rate increases from run R24 to run R22, the radiative cooling gradually strengthens, resulting in the accretion flows gradually contracting.

Figure \ref{Fig4_Density_r} shows the radial profiles of the time-averaged and angle-averaged density near the equatorial plane. Within $r=\sim0.07r_{\rm B}$, we can employ a power-law function $\rho\propto r^{- \rm p}$ to describe the radial profiles of the density near the equatorial plane. In Table \ref{Table1}, we list the values of the power-law index $p$ in a function of $\rho\propto r^{- \rm p}$ for all the runs. As radiative cooling strengthens, the value of $p$ decreases from $0.37$ in run R24 to $0.05$ in run R22. For run N24 without radiative cooling, $p = 0.50$. This indicates that radiative cooling leads to a flatter density distribution in the inner region of the accretion flows. Outside $0.07r_{\rm B}$, the density distribution exhibits different behaviors depending on the specific run. In runs N24 and R24, the density slightly decreases as the flows move inward from $\sim0.3r_{\rm B}$ to $\sim0.2r_{\rm B}$. In the range where $r\sim0.07$--$0.2r_{\rm B}$, a slight increase inward in density is observed. In runs R23 and R22, the density rapidly decreases inward from $0.3r_{\rm B}$ to $\sim0.07r_{\rm B}$. Notably, the decrease inward in density is more pronounced in run R22, which is characterized by stronger cooling effects. This implies that radiative cooling reduces the efficiency of material transfer from a larger radius ($\sim 0.3 r_{\rm B}$) to a smaller radius ($0.1 r_{\rm B}$) becomes low. We will discuss the reason for this phenomenon in Section \ref{subsection_Inflow}.

Figure \ref{Fig5_Beta} shows the time-averaged distribution of the plasma-$\beta$ value and the poloidal mean magnetic field for runs R22 to R24 and N24. The time-averaged plasma-$\beta$ \footnote{For all the runs, turbulence causes plasma-$\beta$ to fluctuate dramatically in both time and space. We found that the time-averaged values of plasma-$\beta$, when calculated using different time-averaging methods, show significant discrepancies. The results obtained using this averaging method are the closest to the statistical characteristics of instantaneous snapshots. We also tested two additional methods for determining the time-averaged values of plasma-$\beta$, such as ${\beta} \equiv {2 \langle p_{g} / \bm{B}^2 \rangle}_{\rm t}$ and ${\beta}\equiv {2 \langle p_{g} \rangle/ \langle \bm{B}^2 \rangle}_{\rm t}$. We found that the results derived from the two averaging methods fail to accurately describe the statistical characteristics of instantaneous states.} is defined as the ratio of the time-averaged gas pressure to the magnetic pressure of the time-averaged magnetic field, i.e., ${\beta} = 2 {\langle p_{\rm g} \rangle}_{\rm t} / {\langle \bm{B} \rangle}_{\rm t}^2$, where ${\langle \bm{B} \rangle}_{\rm t}\equiv  {\langle B_{\rm r} \rangle}_{\rm t} \bm{\hat{r}} + {\langle B_{\rm \theta} \rangle}_{\rm t} \bm{\hat{\theta}} + {\langle B_{\rm \phi} \rangle}_{\rm t} \bm{\hat{\phi}}$ is the time-averaged magnetic field. In run N24 without radiative cooling, the gas pressure around the equatorial plane is stronger than the magnetic pressure, resulting in a relatively weak magnetic field in this region. In runs R24 and R23, which are characterized by weak radiative cooling, similar results are also observed. In contrast, run R22 with strong radiative cooling exhibits a relatively stronger magnetic field around the equatorial plane. In the region near the equatorial plane beyond $\sim2 r_{\rm B}$, the magnetic pressure even exceeds the gas pressure. To further evaluate the relative strengths of the toroidal and poloidal magnetic fields, Figure \ref{Fig6} presents the distribution of the strength ratio between these two types of magnetic fields (${\langle B_{\phi} \rangle}_{\rm t}^2/({\langle B_{\rm r} \rangle}_{\rm t}^2+{\langle B_{\theta} \rangle}_{\rm t}^2)$). In all runs, except for the polar region where the poloidal magnetic field is predominant, the toroidal magnetic field dominates in most regions. Around the equatorial plane, the toroidal magnetic energy accounts for \(60\%\) to \(90\%\) of the total magnetic energy (refer to the dashed line in Figure \ref{Fig20_B_component}).

In the following, we further analyze the reasons for the strengthening of the magnetic fields near the equatorial plane in run R22. The MRI-induced turbulence plays an important role in magnetic field amplification. In run R22, the initial magnetic fields are consistent with those observed in other runs in terms of relative strength and topology. Additionally, the growth time of MRI is also the same in all the runs. It is reasonable to believe that the MRI-induced turbulence leads to a similar effect of magnetic field amplification in all the runs. In addition to the magnetic field amplification resulting from MRI-induced turbulence, the magnetic field strength in accretion flows can be further enhanced by two primary processes: local contraction and field advection by converging flows. For run R22, on the one hand, when radiative cooling weakens the gas pressure support in the accretion flows, the accretion flows undergo significant vertical contraction. On the other hand, magnetic field lines can be advected inward along with the accreting material, transporting magnetic flux from the outer regions to the inner regions and thereby strengthening the field. To quantitatively distinguish between these two mechanisms, we refer to \citet{Tchekhovskoy_2011} and \citet{Narayan_2012} to introduce the normalized magnetic flux:
\begin{equation}
    \dot{\Phi} = \frac{ \Phi  }{\sqrt{\dot{M}_{\rm acc}}},
    \label{mflux}
\end{equation}
where $\Phi = \sqrt{4\pi} {\langle \int_{86\degree}^{94\degree}|B_{\phi}|  r dr \, d\theta \rangle} _t $ is the time-averaged toroidal magnetic flux around the equatorial plane and $\dot{M}_{\rm acc}$ is the net accretion rate (see Table \ref{Table1}). As previously mentioned, the toroidal magnetic field is the primary component near the equatorial plane; therefore, we only consider the magnetic flux of the toroidal magnetic field here. In equation \ref{mflux}, the denominator can reflect the contribution of converging flows to transporting magnetic flux. If only the magnetic flux transport by converging flows is taken into account, both the numerator and denominator in equation \ref{mflux} increase simultaneously. A comparatively small value of $\dot{\Phi}$ indicates that converging flows play a significant role in the transport of magnetic flux. In contrast, a comparatively large value of $\dot{\Phi}$ suggests that the impact of converging flows on magnetic flux transport is relatively minor. Figure \ref{Fig7_MagneticFlux} displays the radial profiles of $\dot{\Phi}$ for all runs. For runs N24, R24 and R23, the radial profiles of $\dot{\Phi}$ are similar. This suggests that the magnetic flux transport by converging flows plays a similar role in strengthening magnetic fields for the three runs. In the region of $r=\sim0.2$--$0.3 r_{\rm B}$, the value of $\dot{\Phi}$ in run R22 is significantly greater than that observed in the other three runs. Comparatively, in run R22, the impact of converging flows on magnetic flux transport is weaker than in the other three runs in the region of $r=\sim0.2$--$0.3 r_{\rm B}$. This region also corresponds to the area with the lowest disk scale height, as shown in Figure \ref{Fig3_DiskH}, and to the region of high density depicted in Figure \ref{Fig4_Density_r}. Therefore, for run R22, it is reasonably inferred that the increase in magnetic field strength is predominantly attributed to the contraction of accretion flows induced by radiative cooling, rather than the magnetic flux transport caused by converging flows.

\subsection{Heating and cooling}

\begin{figure}[htb]
	\includegraphics[width=0.5\textwidth]{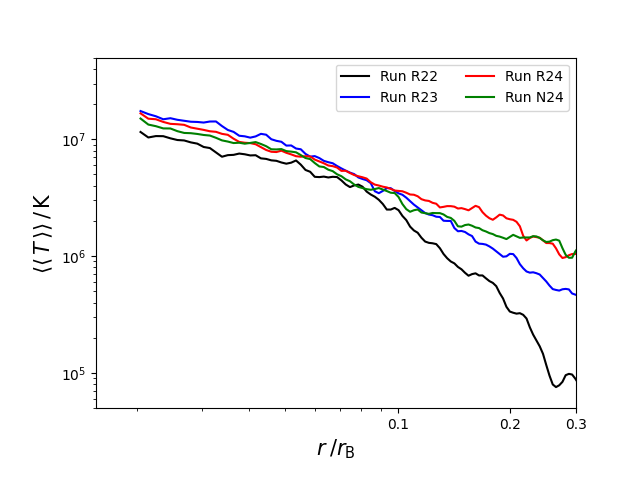}
	\centering \caption{The radial profiles of the time-averaged and angle-averaged temperature near the equatorial plane. The angle averaging is implemented over the angle range of $84\degree$ to $96\degree$. The black, blue, green, and red lines correspond to runs R22, R23, R24, and N24, respectively.}
	\label{Fig8_Tg_r}
\end{figure}

\begin{figure}[htb]
	\includegraphics[width=0.5\textwidth]{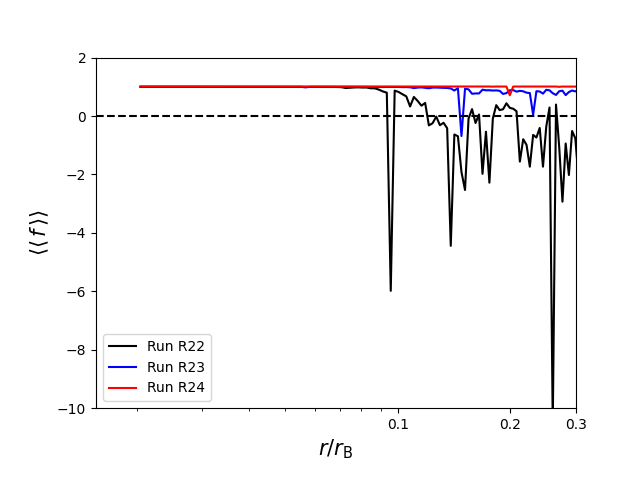}
	\centering \caption{The radial profiles of the time-averaged and angle-averaged advection factor $f$ near the equatorial plane. The angle averaging is implemented over the angle range of $84\degree$ to $96\degree$. The black, blue, green, and red lines correspond to runs R22, R23, R24, and N24, respectively.}
	\label{Fig9_f_r}
\end{figure}

Figure \ref{Fig8_Tg_r} shows the radial profiles of the time- and angle- averaged temperature in the vicinity of the equatorial plane. As shown in Figure \ref{Fig8_Tg_r}, the gas temperature in runs R23, R24 and N24 exhibits similar profiles. The gas temperature in run R23 is slightly lower than that in runs R24 and N24 in the region of $r>0.15 r_{\rm B}$. For run R22, due to strong radiative cooling, the gas temperature is significantly lower than in other runs in the region. The distribution of gas temperature is attributed to cooling and heating. To analyze the energetics of accretion flows, we follow \cite{Yuan_2014} to define the advection factor ($f$) of accretion flows as follows:
\begin{equation}
	f = \frac{Q_{\rm adv}}{Q^{+}} = 1 - \frac{Q^{-}}{Q^{+}}.
	\label{eq_f}
\end{equation}
Here, $Q^{+}$ is the heating rate, $Q^{-}$ is the cooling rate given by equation \ref{eq_Q-}, and $Q_{\rm adv}$ is the advection rate of the gas internal energy.

In magnetic fluids, heating primarily results from the dissipation of magnetic energy. The magnetic energy can be  converted into thermal energy through reconnection in current sheets. In non-conservative numerical schemes, an explicit artificial resistivity term is often incorporated into numerical codes to implement the dissipation of magnetic energy and to effectively capture heating at current sheets (e.g. \cite{Stone_2001}). In conservative numerical schemes, the energy within computational cells can be exchanged between kinetic, magnetic, and thermal forms. In the code of conservative numerical schemes, the inclusion of an artificial resistivity term is unnecessary for capturing heating at current sheets. Numerical simulations in the shearing box have demonstrated that the heating rate equals the work done by the Maxwell and Reynolds stresses at the box boundaries \citep{Hawley_1995, Sano_2004, Gardiner_2005, Jiang_2013}. \citet{Jiang_2014} employed this result to investigate the heating rate in super-Eddington accretion flows. Due to the conservative nature of the numerical schemes employed in this study, here we follow \cite{Jiang_2014} to calculate the heating rate. The local heating rate ($Q^{+}$) is approximately given by
\begin{equation}
	Q^+ = 1.5\Omega_{\rm k} W,
\end{equation}
where $\Omega_{\rm k}$ is the Keplerian angular velocity, and $W$ is the $R\phi$ component of the combined Reynolds and Maxwell stresses in cylindrical coordinates. In spherical coordinates, the stresses are written as
\begin{equation}
	W = -(B_r\sin{\theta} + B_{\theta}\cos{\theta})B_{\phi} + \rho (v_r\sin{\theta} + v_{\theta}\cos{\theta}) \delta v_{\phi},
\end{equation}
where $\delta v_{\phi} = v_{\phi} - \overline{v_{\phi}}$ is the difference between $v_{\phi}$ and the time-averaged $v_{\phi}$.

The advection factor ($f$) defined in Equation \ref{eq_f} is often used to measure the relative importance of advection \citep{Yuan_2014}. Figure \ref{Fig9_f_r} shows the radial profiles of the time-averaged and angle-averaged $f$ near the equatorial plane. In runs R23 and R24, $f\approx 1$, which implies that the advection cooling plays a significant role in the cooling process. For run R22, within 0.1$r_{\rm B}$, $f\approx 1$ and then the advection cooling is a dominant cooling term. Outside 0.1$r_{\rm B}$, $f$ is less than $1$ and even drops to negative values at some radii. This implies that radiative cooling plays a significant role in the energetics outside 0.1$r_{\rm B}$ for run R22, and at some radii the radiative cooling rate exceeds the heating rate.

For the accretion flow near a BH, when $Q_{\rm adv}\approx Q_{+}\gg Q_{-}$ (i.e. $f\approx1$), the accretion flow is radiatively inefficient and then becomes advection-dominated \citep{Narayan_1994, Narayan_1995}. As the accretion rate increases, $f<1$ and then the accretion flow becomes a luminous hot accretion flow \citep{Yuan_2001}. \cite{Bu_2018} simulated the accretion flows with different accretion rates, and found that when radiative cooling is strong, the cooling rate near the equatorial plane exceeds the heating rate at some radii.

\subsection{Convective stability}
\begin{figure}[htb]
    \centering
	\includegraphics[width=0.5\textwidth]{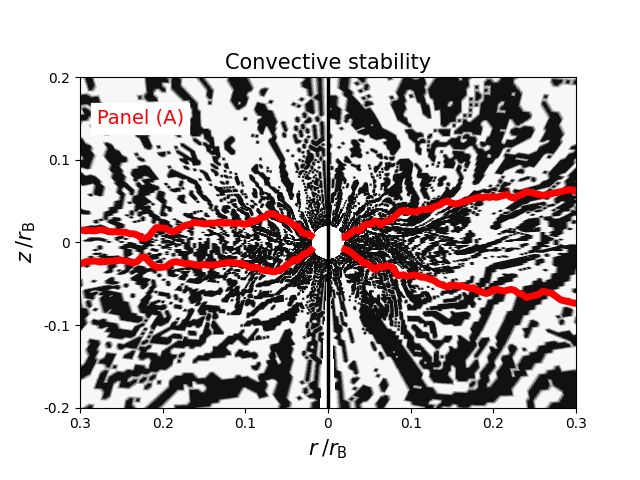}
	\includegraphics[width=0.5\textwidth]{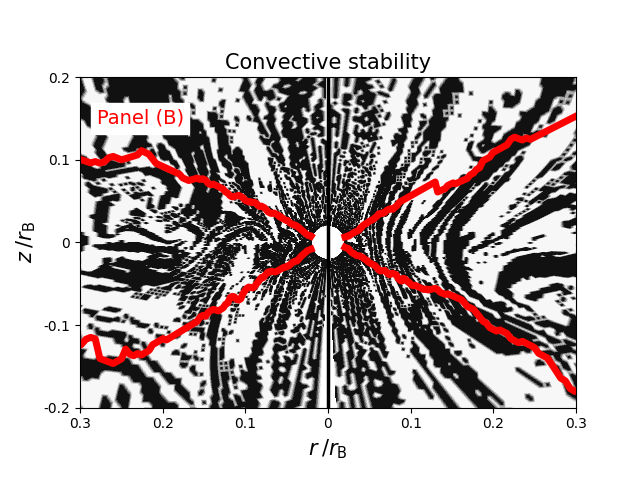}
	
	\caption{Convective stability analysis. The result is obtained according to Equation \ref{eq_CDAF} based on simulation data at $4.5t_{\rm B}$. The red solid lines mark the density scale height of accretion flows, as shown in Figure \ref{Fig2_Density}. Panel (A): Runs R22 (left) and R23 (right). Panel (B): Runs R24 (left) and N24 (right). The black regions are unstable.}
	\label{Fig10_CDAF}
\end{figure}

For a non-rotating flow, an inward increase of entropy can be used as a criterion for convective instability. For a rotating flow, we can employ the H{\o}iland criterion, which is given by \citep{Begelman_1982, Yuan_2012, Bu_2016b}
\begin{equation}
	(\nabla s\cdot\bm{dr})(\bm{g}\cdot\bm{dr})\
	- \frac{2\gamma v_{\phi}}{R^2}[\nabla (v_{\phi}R)\cdot\bm{dr}]dR \textless 0,
	\label{eq_CDAF}
\end{equation}
where $R = r\sin{\theta}$ is the cylindrical radius, $\bm{dr} = dr\hat{r} + rd\theta \hat{\theta}$ is the displacement vector, $s = \ln(p_{\rm g}) - \gamma \ln(\rho)$, and $\bm{g} = -\nabla \Psi + (v_{\phi}^2/R) \hat{R}$ is the effective gravity. When the accretion flow satisfies Equation \ref{eq_CDAF}, it becomes convectively unstable. Figure \ref{Fig10_CDAF} shows the result obtained from Equation \ref{eq_CDAF} based on simulation data at $5 t_{\rm B}$. In Figure \ref{Fig10_CDAF}, the red solid lines mark the density scale height of accretion flows and also indicate the disk surface of accretion flows, as shown in Figure \ref{Fig2_Density}. The black regions are convectively unstable, while the white regions are convectively stable. We have calculated the proportion of the area occupied by the stable regions between the two red lines for each snapshot, and subsequently averaged those proportions during the time interval of 4 to 5 $t_{\rm B}$. Table \ref{Table1} gives the averaged proportion. For runs N24 and R22--R24, the stable regions account  $\sim$ 55 -- 62 $\%$. This implies that the accretion flows around the sub-Bondi radius are marginally stable in convectively stability. For the accretion flows around the BH, \cite{Yuan_2012} employed the H{\o}iland criterion to find that they are convectively stable. In comparison to the accretion flows around the BH, the accretion flows around the sub-Bondi radius promote the occurrence of the convective instability.

\subsection{Outflows}

\begin{figure}[htb]
	\includegraphics[width=0.5\textwidth]{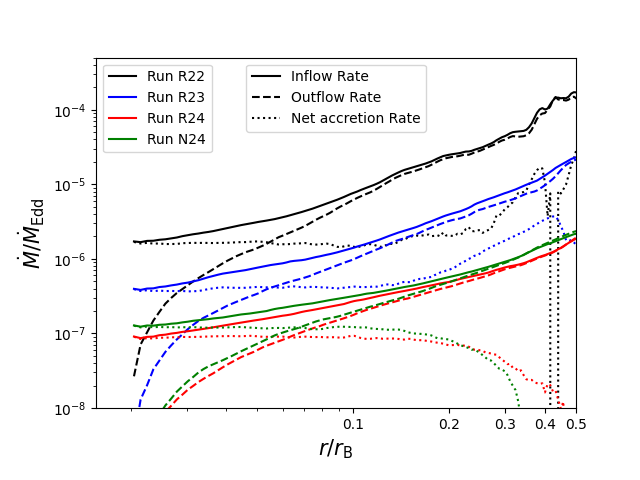}
	\centering \caption{The radial profiles of the time-averaged (from $4 t_{\rm B} $ to $5 t_{\rm B}$ ) mass inflow rates (solid lines), outflow rates (dashed lines) and the net accretion rates(dotted lines). The black, blue, red, and green lines correspond to runs R22, R23, R24, and N24, respectively.}
	\label{Fig11_AccretionRate_r}
\end{figure}

\begin{figure}[htb]
	\includegraphics[width=0.5\textwidth]{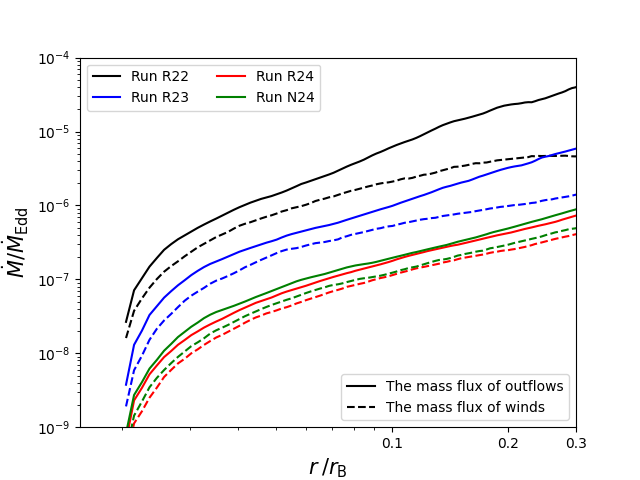}
	\centering \caption{The radial profile of the mass flux of both outflows and winds. Solid lines represent the mass flux of outflows,  while dashed lines denote the mass flux of winds. The black, blue, red, and green lines correspond to runs R22, R23, R24, and N24, respectively.}
	\label{Fig12}
\end{figure}

\begin{figure}[htb]
	\includegraphics[width=0.5\textwidth]{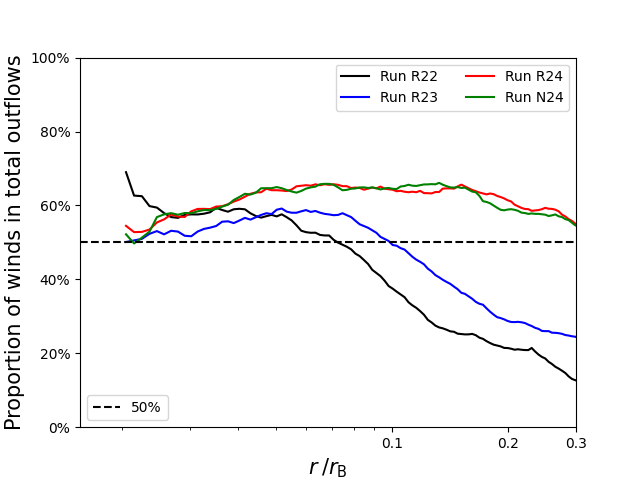}
	\centering \caption{The ratio of the mass flux of the winds (defined as outflows with $Be>0$) to the total mass outflow rate. The black, blue, red, and green lines correspond to runs R22, R23, R24, and N24, respectively.}
	\label{Fig13}
\end{figure}

\begin{figure}[htb]
	\includegraphics[width=0.5\textwidth]{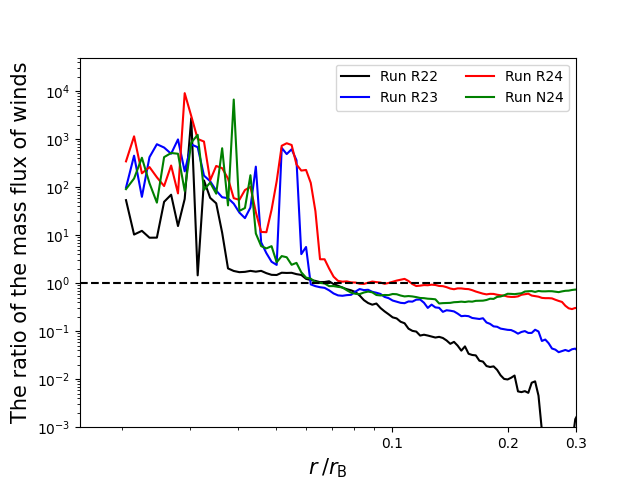}
	\centering \caption{The ratio of the mass flux of winds within disk to the mass flux of winds outside disk. The black, blue, red, and green lines correspond to runs R22, R23, R24, and N24, respectively.}
	\label{Fig14}
\end{figure}

\begin{figure}[htb]
    \centering
	\includegraphics[width=0.5\textwidth]{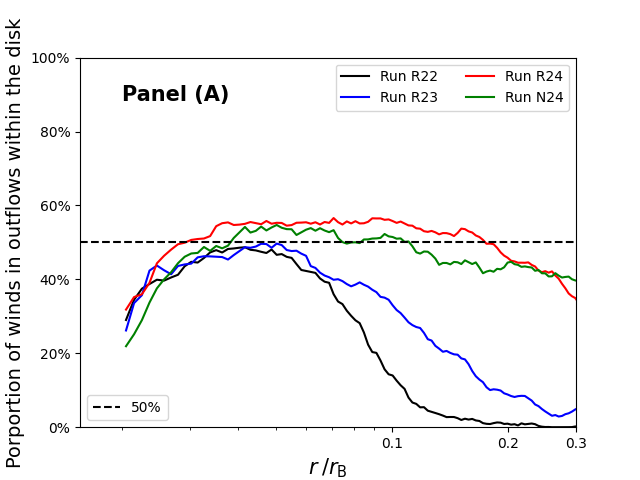}
	\includegraphics[width=0.5\textwidth]{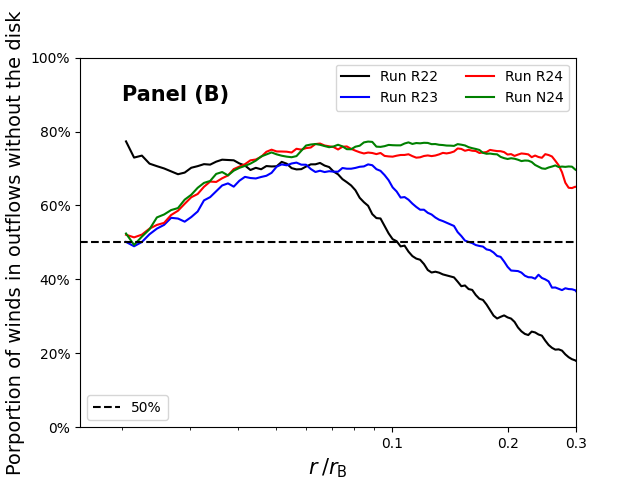}
	\caption{The ratio of winds within and outside the disk to the mass outflow rate. The black, blue, red, and green lines correspond to runs R22, R23, R24, and N24, respectively. Panel (A): the ratio of the mass flux of winds (defined as outflows with $Be>0$) within the disk to the total mass outflow rate within the disk. Panel (B): the ratio of the mass flux of winds outside the disk to the total mass outflow rate outside the disk.}
	\label{Fig15}
\end{figure}

Following \cite{Stone_1999}, we define the mass inflow rate ($\dot{M}_{\rm in}$) and outflow rate ($\dot{M}_{\rm out}$) as:
\begin{equation}
	\dot{M}_{\rm in}(r) = -2\pi r^2 \int_{0}^{\pi}\rho \ \text{min}(v_r,0)\sin{\theta}\ d\theta,
\end{equation}
\begin{equation}
	\dot{M}_{\rm out}(r) = 2\pi r^2 \int_{0}^{\pi}\rho \ \text{max}(v_r,0)\sin{\theta}\ d\theta.
\end{equation}
Then, the net mass accretion rate ($\dot{M}_{\rm net}$) is given by
\begin{equation}
	\dot{M}_{\rm net}(r) = \dot{M}_{\rm in}(r) - \dot{M}_{\rm out}(r).
	\label{Eq_Mdot_acc}
\end{equation}
We first calculate the mass inflow and outflow rates and the net accretion rate from each snapshot and then time-average them.

Figure \ref{Fig11_AccretionRate_r} shows the radial profiles of the mass inflow and outflow rates and the net accretion rate for runs R22--R24 and N24. We can see that the net accretion rate remains approximately constant within $\sim$0.3 $r_{\rm B}$, implying that all the runs achieve a quasi-steady state within this region. As also shown in Figure \ref{Fig11_AccretionRate_r}, the mass inflow and outflow rates decrease inward, similar to the behavior observed in the simulations of accretion flows around the BH \citep{Stone_1999, Yuan_2012}. A power-law function of radius can be employed to approximately describe the radial profiles of $\dot{M}_{\rm in}$ and $\dot{M}_{\rm out}$. We employ the two functions $\dot{M}_{\rm in}(r)\propto r^{s_{\rm in}}$ and $\dot{M}_{\rm out}(r)\propto r^{s_{\rm out}}$ to fit the radial profiles of the inflow and outflow rate, respectively, within the radial range of 0.03--0.3 $r_{\rm B}$. The values obtained from the fitting are given in Table \ref{Table1}. According to Table \ref{Table1}, for runs R23, R24, and N24,  the value of $s_{\rm in}$ ranges from $\sim$0.76 to 1.29, while $s_{\rm out}$ ranges from $\sim$1.34 to 1.68. For run R22, the values of $s_{\rm in}$ and $s_{\rm out}$ are relatively large, being 1.64 and 1.94, respectively. For hot accretion flows near the BH (i.e. at small radii), \cite{Yuan_2012} give $s_{\rm in}\sim0.4$--0.75. In comparison to the the hot accretion flows near the BH, for the accretion flows around the sub-Bondi radius, the mass inflow rate decreases inward more rapidly.

In this paper, the matter moving outward is defined as outflows, while the outward-moving matter with the Bernoulli parameter ($Be$) greater then zero is defined as winds. In magnetic flows, the Bernoulli parameter is given by \cite{Zhu_2018}:
\begin{equation}
	Be = \frac{\bm{v^2}}{2} + \Psi + h + \frac{B_{\phi}B_{\phi}}{\rho} - \frac{B_{\phi}v_{\phi}}{k}.
\end{equation}
Here,  $h = \frac{\gamma p}{(\gamma-1)\rho}$, and $k = \rho \frac{\sqrt{v_r^2 + v_{\theta}^2}}{\sqrt{B_r^2 + B_{\theta}^2}}$. For winds, $Be>0$ implies that they have enough energy to overcome the gravitational potential well and escape to infinity. The inward decrease of the mass inflow rate was attributed to turbulent motion or mass loss caused by winds \citep{Begelman_2012, Narayan_2000}. In the accretion flows with magnetic fields, turbulent motion may be induced by MRI or/and convective instability (\citep{Stone_1999, Stone_2001}). Based on numerical simulations, \cite{Narayan_2012} and \cite{Yuan_2012} found that the hot accretion flows around the BH are convectively stable in the presence of magnetic fields. \cite{Yuan_2015} further found that winds are present in the hot accretion flows and play a key role in the inward decrease of the mass inflow rate. \cite{Bu_2018} found that radiative cooling significantly weakens the winds in the hot accretion flows.

For the hot accretion flows around the \text{sub-Bondi} radius (i.e. all the runs in this work), we plot Figure \ref{Fig12} to compare the radial profile of the mass flux of both outflows and winds. Here, the mass flux of winds is calculated by only considering the contributions from the outflows with $Be >0$. The total mass outflow rate encompasses the contributions from all outflows, including the outflows with $Be >0$ as well as those with $Be\leq0$. In Figure \ref{Fig12}, it is evident that the mass flux of winds decreases as the flows move inward. This trend is consistent with the inward decrease in the mass flux of outflows. We also plot Figure \ref{Fig13} to indicate the ratio of the mass flux of winds to the total mass outflow rate. As shown in Figure \ref{Fig13}, for run N24, the winds account for $50\%$ – $70\%$ of the total mass outflow rate within the quasi-steady state region. For run R22 with strong radiative cooling, the contribution of winds to the mass outflow rate is comparatively lower than in other runs. Specifically, in the region of $r > 0.07 r_{\rm B}$, the contribution of winds falls below 50\%. This implies that the mass outflow rate of winds is significantly reduced due to strong radiative cooling.

In Figure \ref{Fig14}, it is observed that for the region of $r<\sim0.08r_{\rm B}$, the mass flux of winds within the disk is larger than that of winds outside the disk. Conversely, in the region of $r>\sim0.08r_{\rm B}$, the mass flux of winds within the disk is lower than (or comparable to) that of winds outside the disk. For run N24 without radiative cooling, the mass flux of winds within and outside the disk is approximately comparable. As radiative cooling becomes stronger, the winds within the disk is suppressed. In run R22, for the region of $r>\sim0.1r_{\rm B}$, the mass flux of winds within the disk is significantly lower than the winds outside the disk.

To determine the contribution of winds within and outside the disk to the mass outflow rate, we calculated the two ratios. These ratios are presented in Figure \ref{Fig15}. In Figure \ref{Fig15}, the upper panel shows the ratio of the mass flux of winds within the disk to the total mass outflow rate within the disk. The lower panel shows the ratio of the mass flux of winds outside the disk to the total mass outflow rate outside the disk. As shown in the upper panel, for run N24 without radiative cooling, the contribution of the winds within disk account for more than 50 percent of the mass outflow rate within the disk at most radii. As radiative cooling gradually becomes stronger, the contribution of the winds within the disk to the mass outflow rate gradually diminishes. For run R22 with strong radiative cooling, the contribution of the winds within the disk to the mass outflow rate is lower than 40 percent in the region where $r>0.06r_{\rm B}$, and drops below 15 percent for $r>0.1r_{\rm B}$. The lower panel also shows that, for run N24, the contribution of the winds outside the disk account for more than 60 percent of the mass outflow rate outside the disk at most radii. For run R22, outside the disk, the contribution of winds to the mass outflow rate is significantly lower than that in other runs. In run R22, for the region where $r>\sim0.1r_{\rm B}$, the contribution of winds to the mass outflow rate outside the disk decreases to less than 50\%. Therefore, for runs with weak radiative cooling, the mass outflow rate outside the disk is dominated by winds. Additionally, radiative cooling also reduces the mass outflow rate of winds outside the disk, particularly at large radii.

\begin{figure}[htb]
    \centering
	\includegraphics[width=0.5\textwidth]{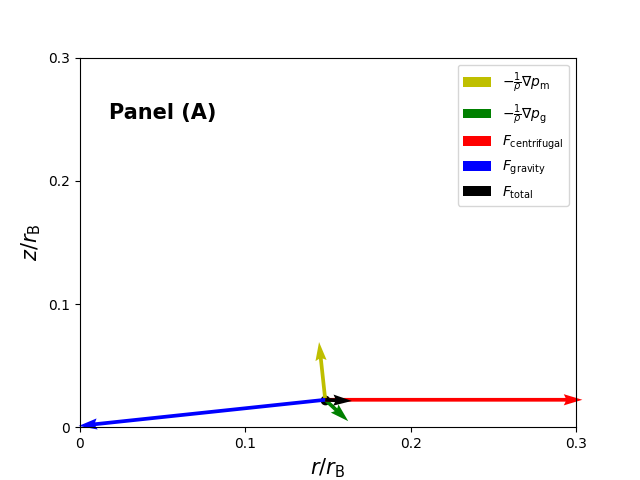}
	\includegraphics[width=0.5\textwidth]{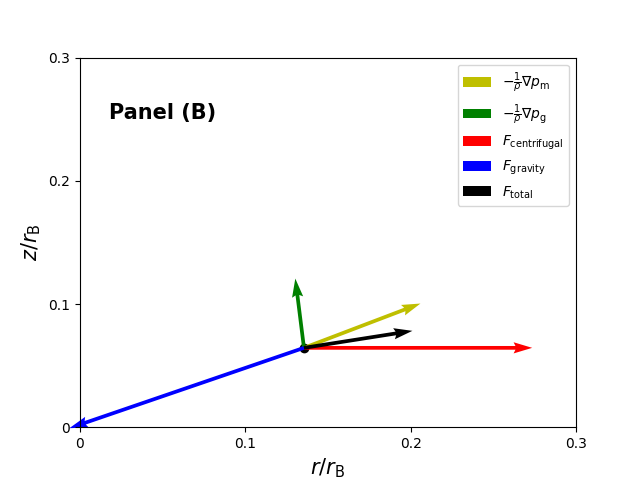}
	\caption{Force analysis at $r = 0.15 r_{\rm B}$ on the disk surface (as indicated by the black line in Figure \ref{Fig2_Density}). The black, blue, red, green, and yellow colors correspond to the net force, gravitational force, centrifugal force, pressure gradient force, and magnetic pressure gradient force, respectively. The arrows point in the direction of each force, and their lengths represent the magnitude of the force. All arrows are scaled identically, with the unit being the magnitude of the gravitational force. Panel (A): Run R22. Panel (B): Run N24.}
	\label{Fig16_Force}
\end{figure}

In addition to the winds that escape to infinity, the outflows that do not reach infinity should be attributed to the outward-moving matter in turbulence. For run N24 without radiative cooling and run R24 with weak radiative cooling, the mass flux of winds exceeds that of the outward-moving matter within turbulent eddies. Therefore, in the two runs, the inward decrease of mass inflow rate is mainly attributed to winds. As radiative cooling gradually becomes stronger, the winds weaken, leading to a reduced influence on the inward decrease of the mass inflow rate, while the role of turbulence becomes increasingly important. For run R22 with strong radiative cooling, in the region where $r<0.05r_{\rm B}$, the contributions of winds and turbulent motions to the inward decrease of mass inflow rate are roughly equal. However, in the region where $r>0.05r_{\rm B}$, the inward decrease of mass inflow rate is mainly attributed to turbulent motions. Panel (A) in Figure \ref{Fig2_Density} shows a snapshot for run R22. As shown in this snapshot, turbulence is significant.

In hot accretion flows, both gas pressure and magnetic forces are commonly considered to play significant roles in the mechanisms that drive winds. \citet{Blandford_1982} suggested the magneto-centrifugally driven mechanism to drive the thin disk winds. Their model necessitates the presence of a large-scale poloidal magnetic field. However, in hot accretion flows, the magnetic field is relatively disordered. \citet{Yuan_2012} also proposed an alternative mechanism for wind production, known as the "micro" Blandford–Payne mechanism. This mechanism indicates that in small-scale turbulent magnetic fields, the combined effects of centrifugal force, magnetic pressure gradient, and gas pressure gradient can drive winds. In the "micro" Blandford-Payne mechanism, the centrifugal force is primarily responsible for accelerating the winds, while the gas pressure gradient force and the magnetic pressure gradient force assist in altering the direction of the outflows or play a comparable role compared with the centrifugal force. To investigate the mechanisms driving winds, we examined the forces within accretion flows in two runs: run N24 without radiative cooling and run R22 with strong radiative cooling. In Figure \ref{Fig16_Force}, we present the time-averaged forces at $r = 0.15 r_{\rm B}$ on the disk surface (as indicated by the black line in Figure \ref{Fig2_Density}) for runs R22 and N24. In the two runs, the centrifugal force plays an important role in driving the winds. In run R22, the net force is directed outward and is weaker than that in run N24. In run N24, the net force is also directed outward and slightly upward. Consequently, the winds in run R22 are significantly weaker than those in the other runs. Therefore, our analysis indicates that the mechanism responsible for driving winds within the accretion flow around the sub-Bondi radius is the "micro" Blandford–Payne mechanism.

\subsection{Inflows}
\label{subsection_Inflow}

\begin{figure}[htb]
	\includegraphics[width=0.5\textwidth]{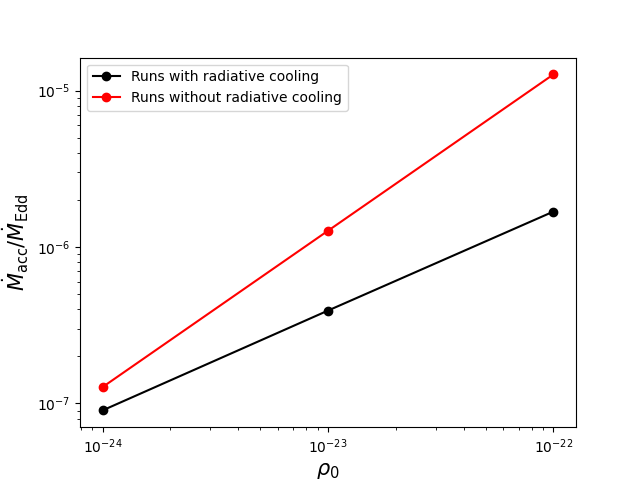}
	\centering \caption{The net accretion rates corresponding to different unit densities, with black and red points representing the cases with radiative cooling switched on and off, respectively.}
	\label{Fig17_Density_Mdot}
\end{figure}

Based on HD simulations, \citet{Li_2013} investigated the effects of radiative cooling on hot accretion flows around the Bondi radius. Their results indicated that as the accretion rate increases, strong radiative cooling promotes accretion. Here, we use MHD simulations to further investigate the effects of radiative cooling on inflows. If the radiative cooling term is not taken into account, the dynamical equations can become density free. Consequently, for the cases without radiative cooling, the results of numerical simulations can be applied to any units of density, and then the accretion rate is directly proportional to the adopted density in simulations. In our simulations, the density of accretion flows is scaled by the density at the center of the initial torus, i.e. $\rho_0$. In Figure \ref{Fig17_Density_Mdot}, we plot the dependence of the net accretion rate on $\rho_0$. The red points correspond to the cases without radiative cooling, whereas the black points correspond to the cases with radiative cooling. For run R24, radiative cooling slightly reduces the net accretion rate due to weak radiative cooling. For run R22 with strong radiative cooling, the net accretion rate is reduced by nearly an order of magnitude compared to the runs without radiative cooling. This suggests that strong radiative cooling suppresses accretion. This result is not consistent with that given by \citet{Li_2013}. In HD simulations, an $\alpha$-viscosity model is adopted to drive angular momentum transport and accretion. In the $\alpha$-viscosity prescription, the effectiveness of angular momentum transport depends on the specific angular momentum flux driven by the anomalous viscosity shear stress. This shear stress is directly proportional to the kinematic viscosity ($\nu$). In the study by \citet{Li_2013}, they assumed a constant, non-zero value for kinematic viscosity ($\nu$). This assumption led to uniform effectiveness in angular momentum transport in all their simulations. In cases of significant radiative cooling, the net accretion rate increases as a result of a reduction in the mass outflow rate caused by the strong cooling effect (refer to Figure 6 in \citealt{Li_2013}). However, in MHD simulations, angular momentum transport is driven by Maxwell and Reynolds stresses. To understand the reduction in the mass accretion rate observed in run R22, we examine the effective viscosity parameter ($\alpha$) and the specific angular momentum flux in MHD simulations. 

\begin{figure}[htb]
	\includegraphics[width=0.5\textwidth]{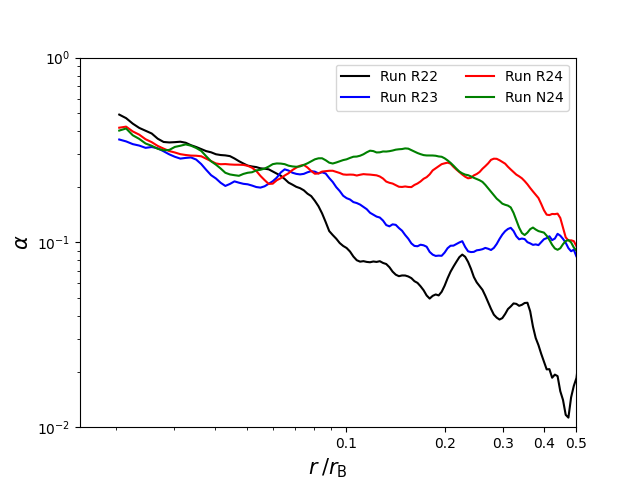}
	\centering \caption{The radial profiles of the effective $\alpha$. The black, blue, red, and green lines correspond to runs R22, R23, R24 and N24, respectively.}
	\label{Fig18_Effective_a}
\end{figure}

In accretion flows, Maxwell stress ($S_{\rm m}$) and Reynolds stress ($S_{\rm r}$) are responsible for the angular momentum transfer. We then follow \citet{Jiang_2019} to calculate the radial profiles of the $r$--$\phi$ stress as $S_{\rm m}=\langle\langle - B_{\rm r}B_{\rm \phi}\rangle\rangle$ and $S_{\rm r}=\langle\langle \rho v_{\rm r}v_{\rm \phi}\rangle\rangle - \langle\langle \rho v_{\rm r}\rangle\rangle \langle\langle v_{\rm \phi}\rangle\rangle$. The effective $\alpha$ can be defined as $\alpha_{\rm m}  \equiv {S_{\rm m}}/{\langle\langle P_{\rm tot}\rangle\rangle}$, $\alpha_{\rm r}  \equiv S_{\rm r}/{\langle\langle P_{\rm tot}\rangle\rangle}$, and then $\alpha \equiv\alpha_{\rm m}+\alpha_{\rm r}$. Here, $P_{\rm tot}$ is the total pressure, which encompasses both the gas pressure and the magnetic pressure. The specific angular momentum flux caused by Maxwell and Reynolds stresses is defined as 
\begin{equation}
    t_{\rm r \rm \phi} = \frac{\alpha r \langle\langle P_{\rm tot}\rangle\rangle}{\langle\langle \rho\rangle\rangle}.
\end{equation}
In Figures \ref{Fig18_Effective_a} and \ref{Fig21_t_rphi}, we plot the radial profiles of the effective $\alpha$ and the specific angular momentum flux, respectively.

\begin{figure}[htb]
	\includegraphics[width=0.5\textwidth]{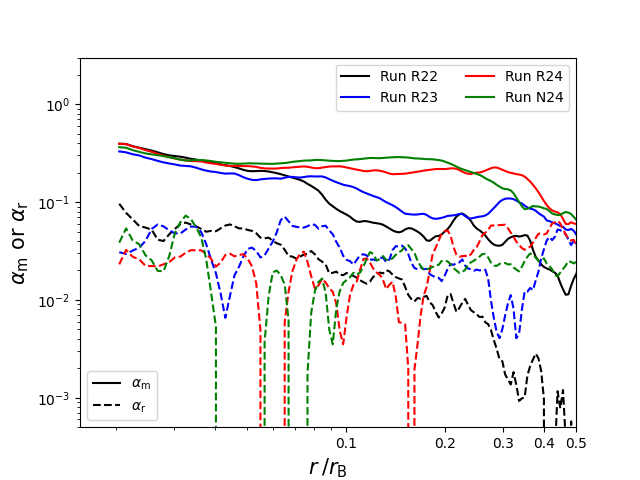}
	\centering \caption{The radial profiles of the effective $\alpha_{\rm m}$ (solid lines) and $\alpha_{\rm r}$  (dashed lines). The black, blue, red, and green lines correspond to runs R22, R23, R24 and N24, respectively.}
	\label{Fig19_a_analysis}
\end{figure}

\begin{figure}[htb]
    \includegraphics[width=0.5\textwidth]{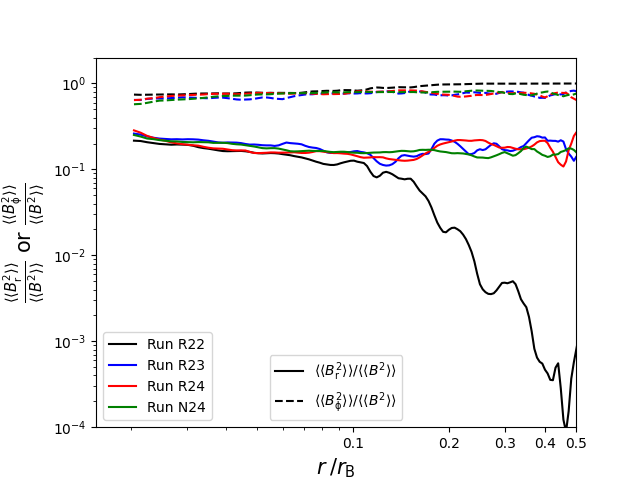}
	\centering \caption{The radial profiles of the percentage of the total magnetic field energy accounted for by the radial (solid lines) and toroidal (dashed lines) components of the magnetic field, respectively. The black, blue, red, and green lines correspond to runs R22, R23, R24 and N24, respectively.}
	\label{Fig20_B_component}
\end{figure}

\begin{figure}[htb]
	\includegraphics[width=0.5\textwidth]{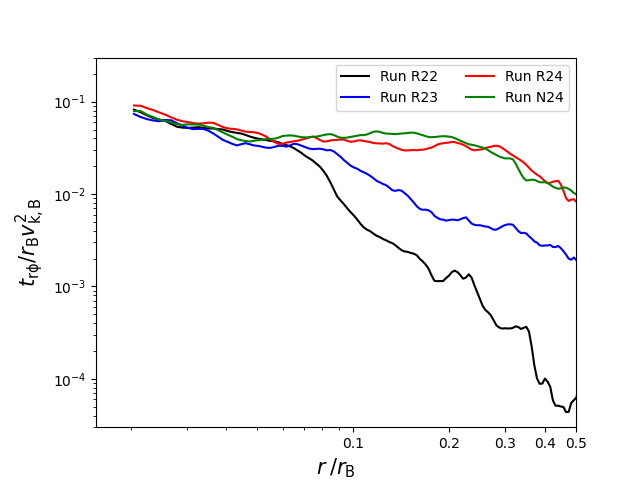}
	\centering \caption{The radial profile of the stress tensor exerted on per unit mass for different runs. The black, blue, red, and green lines correspond to runs R22, R23, R24 and N24, respectively. $r_{\rm B}$ and $v_{\rm k,\rm B}$ denote the Bondi radius and the Keplerian velocity at the Bondi radius, respectively.}
	\label{Fig21_t_rphi}
\end{figure}

\textbf{According to Figure \ref{Fig18_Effective_a}, in the runs (R24 and R23) with weak radiative cooling, the effective $\alpha$ varies from 0.16 to 0.42, with the average value of 0.26. In run R22 with strong radiative cooling, the effective $\alpha$ is higher compared to other runs within the region where $r \lesssim 0.06 r_{\rm B}$. Conversely, for regions where $r > 0.08 r_{\rm B}$, the effective $\alpha$ is lower than those observed in the other runs. In Figure \ref{Fig19_a_analysis}, the effective $\alpha_{\rm m}$ is is significantly greater than the effective $\alpha_{\rm r}$. As a result, the effective $\alpha$ is primarily determined by the effective $\alpha_{\rm m}$. This suggests that Maxwell stresses contribute significantly to the effective $\alpha$. The $r$-$\phi$ component of Maxwell stresses depends on the radial and toroidal components of magnetic fields. As shown in Figure \ref{Fig20_B_component}, the magnetic fields are dominated by the toroidal component. In the region where $r > \sim 0.1 r_{\rm B}$, the radial magnetic field is relatively weak in run R22, leading to a reduction in both the effective $\alpha_{\rm m}$ and $\alpha$ values.}

Figure \ref{Fig21_t_rphi} shows that for run R22 with strong radiative cooling, the specific angular momentum flux driven by Maxwell and Reynolds stresses is significantly reduced in the region of $r > \sim 0.08 r_{\rm B}$. This is because the specific angular momentum flux is influenced not only by the effective $\alpha$ parameter but also by the temperature of accretion flows. In run R22, radiative cooling reduces the temperature of accretion flows (see Figure \ref{Fig8_Tg_r}). A significant reduction in the the specific angular momentum flux at $r=\sim0.4r_{\rm B}$ results in the significant decrease in the mass inflow rate (refer to figure \ref{Fig11_AccretionRate_r}). Consequently, as shown in Figure \ref{Fig4_Density_r}, the radial density profile at the equator in run R22 differs significantly from those observed in other runs with weak radiative cooling. Due to the significant decrease in the mass inflow rate, the net accretion rate in run R22 is significantly reduced compared to the cases with weak radiative cooling (refer to Figure \ref{Fig17_Density_Mdot}). Therefore, accretion in run R22 is inhibited due to significant radiative cooling. 

A low-temperature, high-density accretion disk may become gravitationally unstable, which may result in fragmentation and star formation. Gravitational instability can be assessed using Toomre's criterion \citep{Toomre_1964}:
\begin{equation}
    Q = \frac{C_s \Omega}{\pi G \Sigma},
\end{equation}
where $\Sigma$ is the surface density of the disk. When $Q \geq 1$, the disk is gravitationally stable. When $Q < 1$, the disk is gravitationally unstable. Gravitational stress and Reynolds stress, associated with the gravitational instability-driven turbulence, can provide angular momentum transport in accretion flows. When $Q \ll 1$, the disk is expected to be fragment. We have computed  the $Q$ values and find $Q \gg 1$ for all the runs, which indicates that the disks are gravitationally stable and angular momentum transport is still attributed to Maxwell stresses associated with MHD turbulence driven by the MRI.

\section{Discussions}

For the hot accretion flows near a BH, synchrotron radiation, Compton scattering, and two-temperature effects play significant roles. Nevertheless, these effects are not incorporated into our simulations. The reasons are as follows.

1) When the electron temperature is higher than $\sim10^{10}$\,K, synchrotron radiation becomes important (refer to Figure 1 in \citet{Fragile_2009}). In our simulations, the gas temperature is below $10^{8}$\,K, which is insufficient to produce a significant population of relativistic electrons. Therefore, we do not take into account synchrotron radiation.

2) For the low-luminosity AGNs, Compton scattering plays an important role. The hot accretion flows near a BH can produce a great number of high-energy photons, which can heat the gas at parsec and sub-parsec scales. The significance of this effect depends on the luminosity ($L$) of the hot accretion flows. \citet{Yuan_2011} pointed out that when the luminosity exceeds $2\%$ of Eddington luminosity, the gas at parsec and sub-parsec scales is heated to sufficiently high temperatures. As a result, the gas becomes unbound by the gravity of the central BH, thereby halting the accretion process. After the accretion process ceases, there is a significant decrease in luminosity, rendering Compton heating negligible. Subsequently, the accretion process resumes, leading to intermittent activity. In our simulations, the accretion rate remains consistently low, with a maximum value of $2.5 \times 10^{-6} \dot{M}_{\rm Edd}$. On the other hand, our computational inner boundary is positioned at a considerable distance from the BH. Subsequently, it is unknown how much matter falls into a BH, resulting in difficulty in accurately calculating the luminosity of the hot accretion flows. According to \citet{Wang_2013}, in the matter initially captured by the central BH in Sgr A$^{\star}$, less than 1 \% reaches the innermost region around Sgr A$^{\star}$. Therefore, we believe that for our runs, their luminosity is very low and thus Compton scattering effects are negligible.

3) It is widely accepted that hot accretion flows surrounding a BH exhibit a two-temperature structure, attributed to the weak Coulomb coupling between ions and electrons in the very hot plasma. However, according to \citeauthor{Narayan_1995} \citeyear{Narayan_1995}; refer to Figure 4 in their paper), the temperature gap between ions and electrons diminishes with increasing radial distance and becomes negligible beyond $10^{3} \, r_{\rm s}$. Therefore, in our simulations, two-temperature effects are negligible.

In our simulations, we employ a torus with high angular momentum as the initial condition. Although the specific angular momentum of gases near the Bondi radius ($\sim$11 parsec) is observationally unknown, it is reasonable to infer that the angular momentum of gases in many galaxies is likely not small. Recent kinematic surveys have indicated that approximately 80\% of early-type galaxies (ETGs) are classified as regular rotators \citep{Emsellem_2011}. These galaxies exhibit oblate axisymmetric shapes, which suggest the presence of underlying disk-like components. \citet{Yoon_2018} implemented simulations of elliptical galaxies. In their simulations, the radial range extends from 2.5 parsec to 250 kiloparsec. In simulations of galaxy evolution, gases are introduced through various mechanisms, including AGN mass feedback, stellar mass loss, and supernova explosions. Their results indicate that the gases within a radius of $r<100$ pc have high angular momentum, close to the Keplerian specific angular momentum  (refer to Figure 8 in \citealt{Yoon_2018}). \citet{Gan_2019} further took into account the infall of the circumgalactic medium, and obtained similar results (refer to Figure 6. in \citealt{Gan_2019}).

For all the runs, we consider only the gravitational potential of a non-spinning BH. Since our computational domain is located far from the BH, the effects of BH spin become negligible. However, near the BH, BH spin plays an important role and significantly affect the production and properties of winds. Previous studies have discussed the influence of BH spin on winds in accretion flows \citep{Curd_2023, Aktar_2024}. These winds, launched from the inner region, may propagate outward to the Bondi radius, where they may interact with the interstellar medium on larger scales. Nevertheless, this effect has not yet been investigated in the current study, but it remains an interesting topic for future investigation.

\section{Summary}
The accretion processes surrounding a SMBH in AGNs are thought to span a region extending from the vicinity of the black hole to the Bondi radius. It is difficult to implement numerical simulations on such a large region. Most previous numerical simulations have primarily concentrated on the region close to the BH. These studies have indicated that strong winds can be produced within hot accretion flows near the BH and that radiative cooling plays an important role in modulating the strength of these winds. In this paper, we perform 2D MHD simulations to investigate the impact of radiative cooling on hot accretion flows around the sub-Bondi radius. We exclusively consider radiative cooling through bremsstrahlung radiation. As the mass accretion rate increases, radiative cooling gradually becomes stronger, leading to the contraction of accretion flows and a reduction in the thickness of the accretion disk. In the case of strong radiative cooling, the significant contraction of accretion flows leads to a substantial increase in magnetic pressure within the accretion disk. The radial profiles of the mass inflow and outflow rates can be described by $\dot{M}_{\rm in}(r)\propto r^{s_{\rm in}}$ and $\dot{M}_{\rm out}(r)\propto r^{s_{\rm out}}$, respectively. For the runs with weak cooling, the exponent $s_{\rm in}$ is approximately within the range of 0.7 to 1.3, and the exponent $s_{\rm out}$ is approximately within the range of 1.2 to 1.7. For the run with strong radiative cooling, the inflow and outflow profiles are steeper, with $s_{\rm in} \approx 1.6$ and $s_{\rm out} \approx 1.9$. Similarly, the radial profile of density near the equatorial plane can be expressed as $\rho(r)\propto r^{-p}$. For the runs with weak radiative cooling, the exponent $p$ is approximately within the range of 0.2 to 0.5, while for the runs with strong radiative cooling, the density profile is flatter, with $p \approx 0.05$.

We employ the H{\o}iland criterion to analyze the convective stability of the accretion disk. We find that within the accretion disk (defined as the accretion flows within the density scale height), the region of convection stability accounts for $\sim$55--62\%. This suggests that the hot accretion flows around the sub-Bondi radius are marginally convectively stable.

We define the outflows with $Be>0$ as winds that escape to infinity. In the runs with weak radiative cooling, the winds contribute to $\sim$50--70\% of the mass outflow rate, and then they play a significant role in the inward decrease of the mass inflow rate. In the runs with strong radiative cooling, the mass outflow rate of winds is significantly reduced, particularly at large radii, and then the inward decrease of the mass inflow rate is primarily attributed to turbulence driven by MRI and convection. Radiative cooling suppresses the production of winds.

Furthermore, in the region of $r \gtrsim 0.08 r_B$, the effective viscosity parameter ($\alpha$) in the runs with strong radiative cooling is significantly lower than in those with weak radiative cooling. As a result, accretion processes are suppressed under conditions of strong radiative cooling. This result differs from the results obtained from HD simulations.

\begin{acknowledgements}
We thank the anonymous referee for the valuable suggestions, which were very helpful in improving the manuscript. This work is supported by the Natural Science Foundation of China (grant 12347101). D. Bu is supported by the Natural Science Foundation of China (grant 12192220, 12192223). 
\end{acknowledgements}

\bibliographystyle{aa}
\bibliography{ref}

\begin{appendix}
\nolinenumbers
\section{Assessing the outer boundary sensitivity and the impact of simulation dimensionality}
\begin{table}
\centering
\caption[\label{TableA1}]{Main parameters of additional simulation test runs}
\begin{tabular}{l c c c}
\toprule[1.3pt]
Run name &  $\dot{M}_{\rm acc}(\dot{M}_{\rm Edd})$  & $r_{\rm out}(r_{\rm B})$ & Dimensional\\
\midrule[1.1pt]
N24a   & $2.0 \times 10^{-7}$  & $5$  & Two   \\
N24b  & $2.2 \times 10^{-7}$  & $10$ & Two   \\
N24c  & $2.2 \times 10^{-7}$  & $5$  & Three \\
\bottomrule[1.3pt]
\end{tabular}
\begin{tablenotes}{}

\item{Notes. Column (1) means run name. 
Column (2) gives the net accretion rate at inner boundary. 
Column (3) gives the outer boundary radius in units of the Bondi radius. 
Column (4) gives the dimensionality of the simulation. }
\end{tablenotes}
\end{table}

\begin{figure}[htb]
	\includegraphics[width=0.5\textwidth]{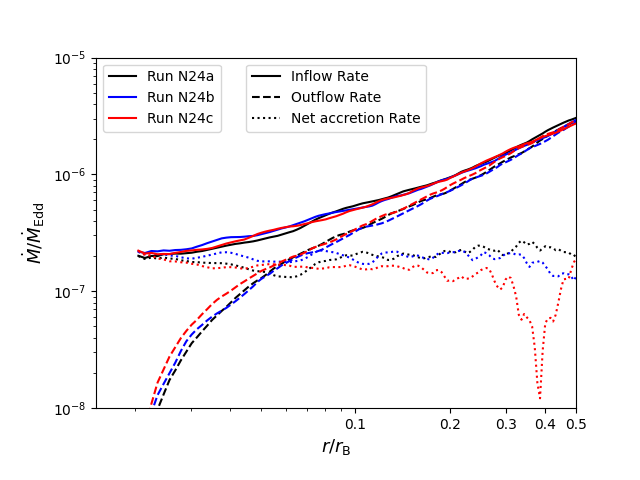}
	\centering \caption{The radial profiles of the time-averaged mass inflow rates (solid lines), outflow rates(dashed lines) and net accretion rates(dotted lines). The black, blue and red lines correspond to runs N24a, N24b and N24c, respectively.}
	\label{FigA1_AccretionRate_Compare}
\end{figure}

\begin{figure}[htb]
	\includegraphics[width=0.5\textwidth]{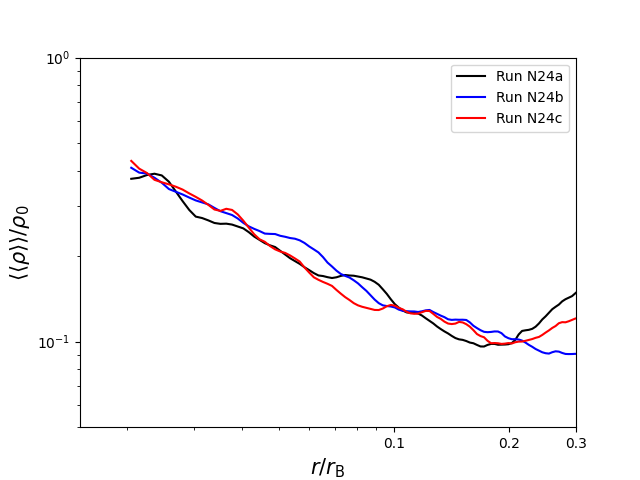}
	\centering \caption{The radial profiles of the time-averaged and angle-averaged density near the equatorial plane. The angle averaging is implemented over the angle range of $84\degree$ to $96\degree$. The black, blue and red lines correspond to runs N24a, N24b and N24c, respectively.}
	\label{FigA2_Density_Compare}
\end{figure}

We implemented run N24b to assess the sensitivity to the outer radius of the computational region, and run N24c to evaluate the impact of simulation dimensionality. In the following analysis, we perform time-averaging over the time interval of 2--4 $t_{\rm B}$. During this time interval, we gather 200 data files with equal time intervals. Run N24b is an alternative version of run N24a. Compared to run N24a, run N24b has a larger outer boundary. The outer boundary of run N24b is set at $10 r_{\rm B}$. In run N24b, the grid resolution is consistent with that of Run N24a within $r<5r_{\rm B}$. In the radial range extending from $r = 5r_{\rm B}$ to $10 r_{\rm B}$, we employ 68 uniformly distributed zones. Run N24c is a 3D version of run N24a. In the $r$ and $\theta$ directions, runs N24c and N24a have the same number of zones. In the $\phi$ direction, we employ 64 uniformly-spaced zones for run N24c. Compared to run N24a, runs N24b and N24c exhibit qualitatively consistent properties, as evidenced by the data presented in Table A.1. Figure \ref{FigA1_AccretionRate_Compare} shows the radial profiles of the net accretion rate, inflow rate, and outflow rate for comparing runs N24a (black lines) and N24b (blue lines). Figure \ref{FigA2_Density_Compare} also presents the radial density profiles at the equatorial plane. Figures \ref{FigA1_AccretionRate_Compare} and \ref{FigA2_Density_Compare} demonstrate that the radial profiles of these quantities are comparable in runs N24a, N24b and N24c. This suggests that our simulation results are not sensitive to the position of the outer boundary, provided the outer boundary is situated at a substantial distance from the initial torus, and also indicates that the axisymmetric 2D simulations are sufficient to capture the key features of the accretion flow dynamics in our study.

\end{appendix}

\end{document}